\documentclass[a4paper]{article}

\usepackage{jheppub} 

\usepackage{graphicx,color}
\usepackage[export]{adjustbox}
\usepackage{bm}
\usepackage{amsmath}
\usepackage{amssymb}    
\usepackage{pifont}
\usepackage{stmaryrd}
\usepackage{physics}    
\usepackage{slashed}
\usepackage{appendix}
\usepackage{caption}
\usepackage{subcaption}
\usepackage{float}

\usepackage{standalone}
\usepackage{multirow}
\usepackage{makecell}
\usepackage{boldline}
\usepackage{tabu}
\usepackage{hhline}
\usepackage{colortbl}
\usepackage{gensymb}

\usepackage[utf8]{inputenc}
\usepackage[T1]{fontenc}
\usepackage{amsfonts}
\usepackage[version=4]{mhchem}
\usepackage{mathrsfs}

\newcommand{\nn}{\nonumber}
\newcommand{\df}{\mathrm{d}}

\newcommand{\bmat}[1]{\boldsymbol{#1}}

\renewcommand{\(}{\left(}
\renewcommand{\)}{\right)}
\renewcommand{\[}{\left[}

\newcommand{\img}{\mathrm{i}}

\title{
Event-axis TMD measurements in $e^+e^-$ and SIDIS}

\newcommand*{\UCM}{Departamento de F\'isica Te\'orica \& IPARCOS, Facultad de Ciencias Físicas, Universidad Complutense de Madrid, Plaza de Ciencias 1, E-28040 Madrid, Spain}
\newcommand*{\UvA}{Institute of Physics, University of Amsterdam, Science Park 904, Amsterdam, 1098 XH, The Netherlands}
\newcommand*{\Nikhef}{Nikhef, Theory Group, Science Park 105, 1098 XG, Amsterdam, The Netherlands}

\author{Daniel Díaz Fernández$^{a}$,}
\emailAdd{danidi03@ucm.es}
\author{Patricia Andrea Gutiérrez García$^{a}$,}
\emailAdd{patricgu@ucm.es}
\author{Ignazio Scimemi$^{a}$,}
\emailAdd{ignazios@ucm.es}
\author{Wouter Waalewijn$^{b,c}$}
\emailAdd{w.j.waalewijn@uva.nl}
\affiliation{$^a$\UCM}
\affiliation{$^b$\UvA}
\affiliation{$^c$\Nikhef}

\preprint{IPARCOS-UCM-26-30}

\abstract{
Transverse-momentum-dependent (TMD) fragmentation in $e^+e^-$ collisions can be studied by measuring hadrons with respect to the thrust axis, and  has been measured at Belle. This provides a complementary way to extract TMD fragmentation functions, avoiding the need to disentangle the two TMD fragmentation functions that enter conventional back-to-back hadron-pair measurements.
Starting from the established factorization theorems for this observable, we complete the operator-level formulation of the soft ingredients and perform one-loop checks using the $\delta$-regulator.
We also extend existing results for 1-jettiness factorization in semi-inclusive deep-inelastic scattering (SIDIS), where analogous measurements give access to the TMD parton distribution functions of the incoming hadron.
For phenomenology, we discuss the nonperturbative effects and propose a model that captures both the event-shape dependence and correlations between the event-shape and transverse-momentum measurements.
We resum the transverse-momentum and thrust logarithms, explore several schemes for treating the latter, and implement it in \texttt{artemide}.
As a first validation, we compare to simulated $e^+e^-$ data from \textsc{Pythia}~8.3.
We find that the proposed nonperturbative model is flexible enough to describe the simulated data, with fitted parameters of the expected size in powers of $\Lambda_{\rm QCD}/Q$.
In this test, the resummation of the logarithms of $q_T/(\tau Q)$ appears to have little impact on the fit quality, but changes the fit parameters.
}

\begin{document} 
\allowdisplaybreaks
\maketitle

\begin{figure}
  \hspace{-2cm}
    \includegraphics[width=1.2\linewidth]{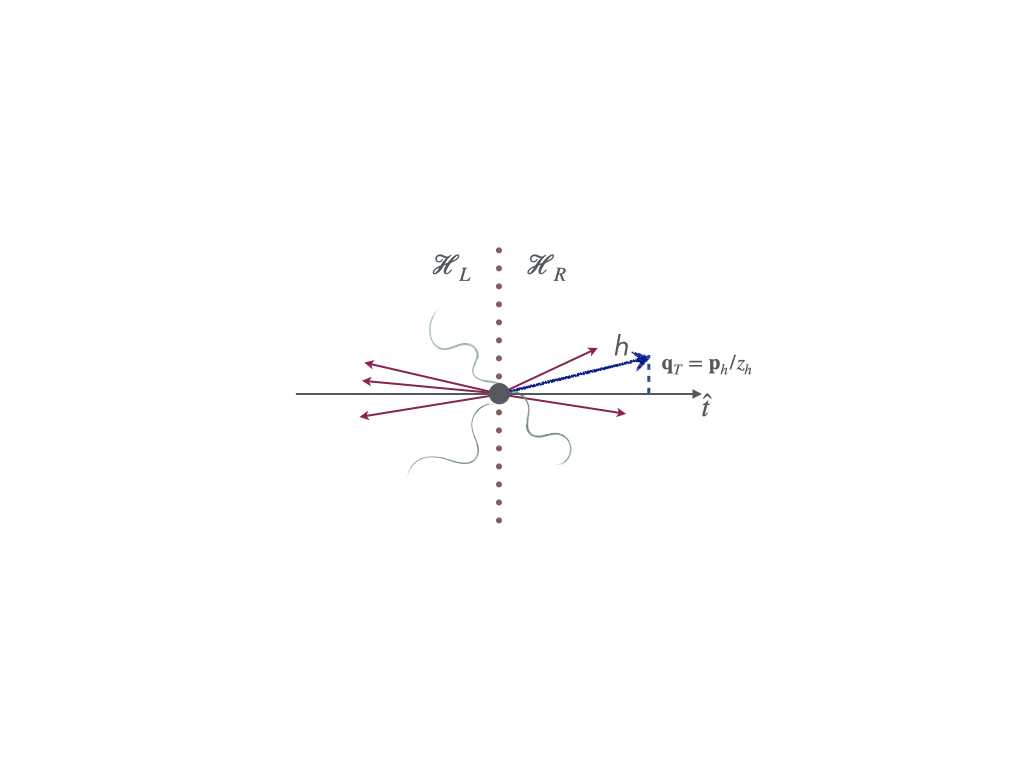}
   \vskip-4cm
    \caption{ The thrust event shape separates the phase space into two hemispheres $\mathcal{H}_{L/R}$. For small values of thrust $\tau$, we have two jets of collinear radiation (red) and softer wide angle radiation (green). The transverse momentum $q_T$ of a fragmenting hadron (blue line) with respect to the thrust axis $\hat t$ is shown.
    }
    \label{fig:ThrustHemis}
\end{figure}

\section{Introduction}

The measurement of transverse-momentum-dependent distributions (TMDs) in both initial and final 
hadronic states plays a central role in the physics programs of  current and future experiments, including the Electron–Ion Collider (EIC)~\cite{AbdulKhalek:2021gbh}, JLab~\cite{Accardi:2023chb,Abir:2023fpo}, AMBER~\cite{Quintans:2022utc,Ketzer:2026ytc}, and LHCspin~\cite{LHCspin:2025lvj} at CERN, as well as Belle~\cite{Belle:2008fdv,Belle:2011cur,Belle:2018ttu,Belle:2022fvl,Belle:2022ars}.  These distributions are referred to as TMD parton distribution functions (TMDPDFs) and TMD fragmentation functions (TMDFFs), and enter the factorization theorems for processes such as Drell–Yan, semi-inclusive deep inelastic scattering (SIDIS), and $e^+e^-$ annihilation to two hadrons (semi-inclusive annihilation or SIA) or jets. 
In factorization theorems the TMDs are accompanied by the Collins-Soper (CS) evolution kernel, which also has a nonperturbative part.
The TMDs are process independent, i.e.~universal, enabling global analyses. 
Extractions from data and results from nonperturbative methods indicate an approximate convergence  for the latter~\cite{Bertone:2019nxa,Scimemi:2019cmh,Moos:2023yfa, Moos:2025sal,Bacchetta:2022awv,Bacchetta:2024qre,Bacchetta:2025ara,
Bollweg:2024zet,Avkhadiev:2023poz, Avkhadiev:2024mgd,Shu:2023cot,LatticePartonLPC:2022eev,Kang:2024dja, Cuerpo:2025zde,Aglietti:2026bhu,Liu:2024sqj,Aslan:2024nqg,
Boglione:2023duo,Konychev:2005iy, Ebert:2018gzl, Shanahan:2020zxr}. 
On the other hand, for TMDPDFs, and especially TMDFFs, there is not a consensus on their extraction, see e.g.~\cite{Bacchetta:2022awv,Moos:2025sal,Bacchetta:2025ara}.
 Exploring different methods for measuring TMDs can therefore be crucial to better understand these distributions. 

In this work we study the theoretical and phenomenological framework required to analyze TMDs with respect to an event axis, including higher-order perturbative contributions. In $e^+e^-$ collisions the thrust axis provides a natural reference direction for measuring TMDFFs,  which has been used to measure TMD fragmentation at Belle~\cite{Belle:2019ywy}.
Thrust is defined as~\cite{Farhi:1977sg},
\begin{equation} \label{eq:thrust}
    1-\tau=T= \max _{\hat{t}} \frac{\sum_{i}\left|\hat{t} \cdot \vec{p}_{i}\right|}{\sum_{i}\left|\vec{p}_{i}\right|}\,,
\end{equation}
where the sum runs over all particles $i$ in the final state with 3-momentum $\vec p_i$. The unit vector vector $\hat t$ that maximizes $T$, defines the direction of the thrust axis, and naturally divides the final state into two hemispheres depending on the sign of $\hat{t} \cdot \vec{p}_{i}$, see fig.~\ref{fig:ThrustHemis}.  The maximum value of $\tau=1/2$ corresponds to spherically symmetric events. Experimentally most events reported in ref.~\cite{Belle:2019ywy} have $\tau \ll 1$, corresponding to two highly-collimated back-to-back jets. The production of a hadron in this process is illustrated in fig.~\ref{fig:ThrustHemis}, with
\begin{equation} \label{eq:meas}
    z_h=\frac{2E_h}{Q}\,,\quad p_{hT} \equiv |\boldsymbol{p}_{h}| \,, \quad q_T=\frac{p_{hT}}{z_h}\,,
\end{equation}
where $E_h$ is the hadron energy,  and $p_{hT}$ is its transverse momentum with respect to the thrust axis.

TMD distributions are present in the leading power factorization of the cross section
$$\frac{{\rm d}\sigma}{{\rm d} z_h\,{\rm d}q_T\,{\rm d}\tau}$$ in $e^+e^-$ 
annihilation, as has been established in ref.~\cite{Makris:2020ltr}. We complete the factorization analysis providing the missing operator definitions. We perform a one loop check, regulating rapidity divergences using the $\delta$-regulator and discussing the zero-bin subtraction. We describe all perturbative contributions that we need to achieve next-to-next-to-leading logarithmic order, including the two loop anomalous dimension of the collinear-soft function, obtained from consistency. We also discuss all nonperturbative inputs necessary for a full description of the process,  including a new nonperturbative model that accounts for the thrust measurement and possible correlations with the TMD measurement.
Because the zero-bin subtraction is provided  by the same soft function that appears in Drell-Yan and SIDIS, the distributions are universal and  can be used in global fits.

We perform a preliminary validation of our framework by comparing our predictions with simulations from \textsc{Pythia}~8.3~\cite{Bierlich:2022pfr}.  The purpose is to test the nonperturbative structures that depend on thrust  and the resummation procedure. Fitting simulated \textsc{Pythia} data points, we discuss several resummation schemes. We first consider two schemes where logarithms of both $\tau Q/q_T$ and $q_T/Q$ are resummed  (dubbed $\tau$-scheme and $u$-scheme). Then we consider a scheme where only the logarithms of   $q_T/Q$ are resummed (the $\mathcal{M}$-scheme). 
More in detail, we implement the resummation using the so-called $\zeta$-prescription~\cite{Scimemi:2018xaf}  in the \texttt{artemide}~\cite{artemide} code.
The fits result in similar $\chi^2/N$ for the different schemes, however the value of the nonperturbative parameters and the interpretation of the result changes.

\begin{figure}
    \centering
    \includegraphics[width=0.45\linewidth]{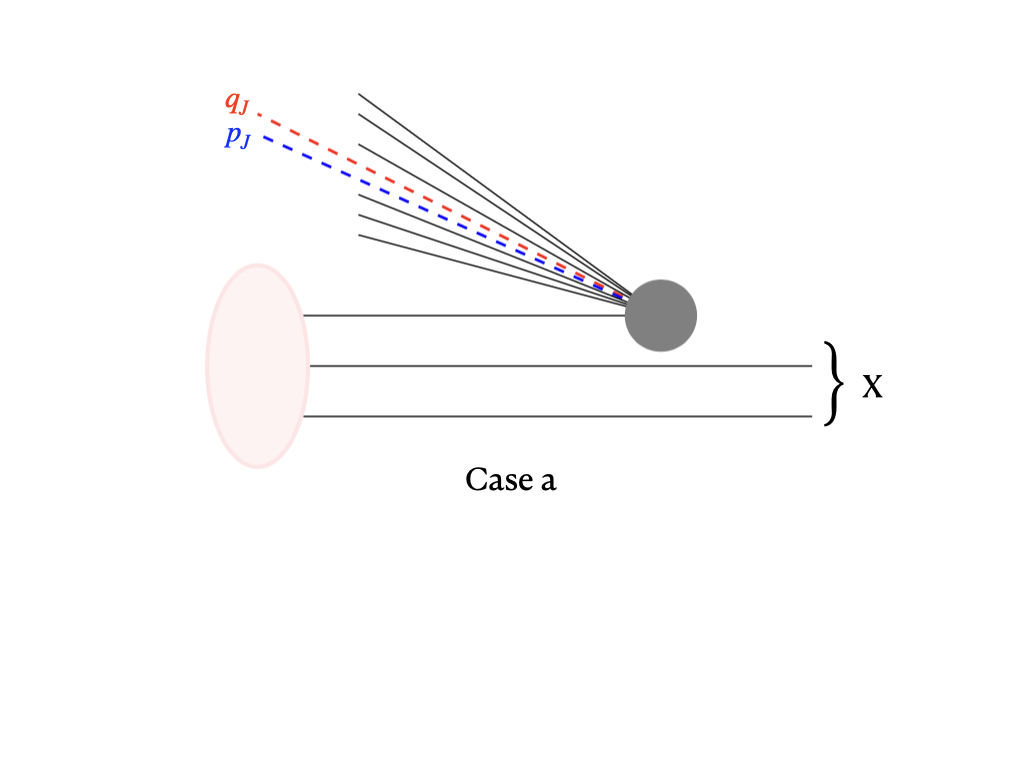}
      \includegraphics[width=0.45\linewidth]{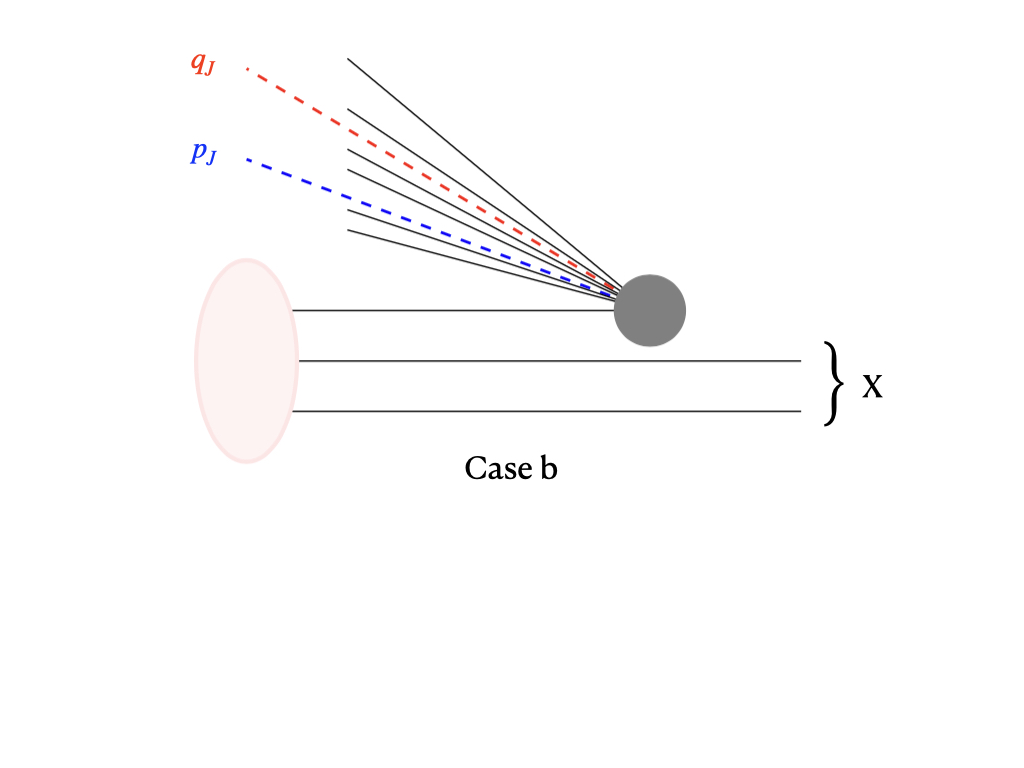}
   \vskip-1cm
    \caption{SIDIS in the Breit frame. The two cases correspond to whether the 1-jettiness axis ($q_J$) is aligned (case a) with the jet momentum $p_J$ or not (case b).}
    \label{fig:SIDIScases}
\end{figure}

The factorization analysis for $e^+e^-$ can be extended to SIDIS through 1-jettiness.
Ref.~\cite{Kang:2013nha}  analyzed this event shape in inclusive deep inelastic scattering (DIS). We extend their factorization analysis by including the measurement of a transverse momentum 
in the final state leading to a TMDPDF in the initial state.
 We consider two possibilities for the 1-jettiness axis, depending on whether the jet momentum is aligned with it or not, see fig.~\ref{fig:SIDIScases}.
 In case of alignment (case $a$)  the factorization is obtained for the measurement of the transverse momentum of the mediating vector boson.
 Otherwise (case $b$), the transverse momentum 
of the jet with respect to the 1-jettiness axis is still relatively small, and the factorization is obtained for the transverse momentum of the jet.
These cases are relevant because they show how  different TMD measurements are possible using event shapes in SIDIS. The nonperturbative  inputs depending on thrust and present in the resummation in SIDIS and $e^+e^-$ are the same, which opens the path to a global treatment in data analysis. 

The paper contains a detailed description of the factorization for TMD fragmentation with respect to the thrust axis in $e^+e^-$ collisions in sec.~\ref{sec:SIA}, which includes the operator definition and one-loop calculation of all soft ingredients. In sec.~\ref{sec:SIDIS} the factorization for SIDIS is presented.
The resummation is discussed in sec.~\ref{sec:resummation} and the nonperturbative ingredients in sec.~\ref{sec:Models}.
Our results of the comparison with \textsc{Pythia}~8.3 are presented in sec.~\ref{sec:results}
after which we present our conclusions in sec.~\ref{sec:conclusions}.

\section{Factorization for $e^+e^-$}
\label{sec:SIA}
In this section we present the kinematics and factorization of the cross section for the process  $e^+e^-\rightarrow h X$ with a cut on the thrust event shape $\tau$. The cross section will be differential in the hadron momentum fraction $z_h$ and transverse momentum, $\boldsymbol{q}=\boldsymbol{p}_h/z_h$. As usual, the vector boson momentum is indicated by $q^\mu$ and $Q^2=q^2$.

It will be convenient to decompose momenta into light-cone coordinates, $p^\mu = (p^+,p^-, \boldsymbol{p})$, for which we adopt the following convention
\begin{align}
 p^+ &= n\cdot p,\;\quad  p^- = \bar{n}\cdot p,\;\quad    p^2 = 2p^+ p^- - \boldsymbol{p}^{\, 2} \,,\quad
 p_T=|\boldsymbol{p}|\,,
\end{align}
with $n^\mu=(1,0,0,1)/\sqrt{2}$, $\bar n^\mu=(1,0,0,-1)/\sqrt{2}$.

The thrust event shape and hadron transverse momentum can be written as
\begin{align}
    \tau=\frac{\sqrt{2}}{Q}\sum_i\text{min}\{p_i^+,p_i^-\}\,, \qquad \boldsymbol{p}_h=-\sum_{\underset{i\neq h}{i\in\text{hemi}}} \boldsymbol{p}_{i}\,.
\end{align}
The expression for $\tau$ assumes that the thrust axis coincides with the $z$-axis and that all particles are massless. The plane perpendicular to the thrust axis divides the phase space into two hemispheres, corresponding to a positive or negative momentum along the $z$-axis, see fig.~\ref{fig:ThrustHemis}. The expression for the hadron transverse momentum relies on a property of the thrust axis: the total transverse momentum with respect to the thrust axis is zero in each hemisphere.

We will now discuss the various kinematic regions, treating $z_h$ as order 1 throughout. When $q_T/Q \sim 1$ the transverse momentum of the hadron is as large as the hard scale and is generated by the hard scattering. In this case, the only nonperturbative information is encoded in the
collinear fragmentation functions that depend on $z_h$, up to corrections of $\mathcal{O}(\Lambda_{\rm QCD}^2/Q^2)$.  We will instead assume $q_T\ll Q$, where more nonperturbative information is needed to account for $\mathcal{O}(\Lambda_{\rm QCD}^2/q_T^2)$ effects.
Due to the additional measurement (or cut) on $\tau$, one identifies the following three kinematic regions~\cite{Procura:2014cba, Makris:2020ltr}
\begin{align}
  \text{Region 1:}\quad &\tau \sim q_T/Q \ll \sqrt{\tau}\,, \nn\\ 
  \text{Region 2:}\quad&\tau \ll q_T/Q \ll \sqrt{\tau}\,,\nn \\ 
  \text{Region 3:}\quad&\tau \ll q_T/Q \sim \sqrt{\tau}\,. 
  \label{eq:regions}
\end{align}
Soft radiation always contributes to thrust, and collinear radiation in the hemisphere with the observed hadron always affects the measured transverse momentum. Whether this radiation is resolved by both measurements depends on the relative size of $q_T$ and $\tau$, giving rise to these three regions.
The region $q_T/Q \gg \sqrt{\tau}$ is not kinematically possible, while $q_T/Q \ll \tau$ is possible but corresponds to the suppressed situation with a large hierarchy in radiation between the two hemispheres, requiring the resummation of non-global logarithms~\cite{Dasgupta:2001sh}.
We will consider $q_T \gtrsim \Lambda_\text{QCD}$, such that the lowest admissible values for (a cut on) $\tau$ are $\tau\gtrsim \Lambda_\text{QCD}/Q$.

The modes in Soft-Collinear Effective Theory for these different regions are summarized in table~\ref{tab:modes}. In the next subsections we will discuss the factorization for all three regions, though our main focus for phenomenology will be on region 2.
We will also present the definitions of the operators entering in these factorizations, except for the (unpolarized) TMD fragmentation function which is well known in the TMD literature and whose matching onto collinear fragmentation functions has been calculated to higher order in perturbation theory~\cite{Echevarria:2016scs,Luo:2019hmp,Ebert:2020qef}.

In the following we will use the Fourier transform and Laplace transform of a function $f$, as well as their inverse. Our conventions are summarized below: 
\begin{align} \label{eq:FT_LT}
  f(\bmat{b}) &= \int\! {\rm d} \bmat{q}\; e^{-\img \bmat{b}\cdot \bmat{q}} \, f(\bmat{q})\;, &
  f(\bmat{q}) &= \int\!  \frac{{\rm d} \bmat{b}}{(2\pi)^2}\; e^{\img \bmat{b}\cdot \bmat{q}} {f}(\bmat{b}) \;,
 \nn \\
  f(u) &= \int_{0}^{\infty}\! \df \tau\; e^{-u\tau}\, f(\tau) \;, &
  f(\tau) &= \frac{1}{2\pi i} \int_{\gamma-i \infty}^{\gamma+i \infty}\! \df u\; e^{u\tau}\, f(u) \;. 
\end{align}

\begin{table}[t]
\renewcommand{\arraystretch}{1.5}
\centering
   \begin{tabular}{|l|ccc|} 
     \hline \hline
     Regime: & 1: $\tau \sim q_T/Q \ll \sqrt{\tau}$ & 2: $\tau \ll q_T/Q \ll \sqrt{\tau} $ & 3: $ \tau \ll q_T/Q \sim \sqrt{\tau}$ \\
     \hline
     $n$-collinear & $(\lambda^2 ,1,\lambda)$  & $(\lambda^2 ,1,\lambda)$ & $(\tau,1,\sqrt{\tau})$  \\
     $\bar n$-collinear & $(1,\tau, \sqrt{\tau})$ & $(1,\tau, \sqrt{\tau})$ & $(1,\tau,\sqrt{\tau})$ \\
     $n$-collinear-soft & & $(\tau ,  \lambda^2 /\tau, \lambda)$ & \\
     soft & $\lambda (1,1,1)$ & $\tau (1,1,1)$ & $\tau (1,1,1)$ \\
     \hline \hline
   \end{tabular} 
   \caption{Modes (rows) and the corresponding power counting of light-cone components $(p^+, p^-, p_T)$ of momenta for the three kinematic regions (columns). Here $\lambda\sim q_T/Q$, and we have omitted a common factor $Q$ in front of all momenta. There is only a $n$-collinear-soft mode, because we assume that the measured hadron is $n$-collinear.}
    \label{tab:modes}
\end{table}

\subsection{Factorization of the cross section in $e^+e^-$}

We now present the factorization for the various kinematic regions, which was established in ref.~\cite{Makris:2020ltr}.
The cross section for region 3 ($\tau Q \ll q_{T} \sim \sqrt\tau Q$) is given by
\begin{align}
\label{eq:Xsec3i}
\frac{\mathrm{d}\sigma_{3}}{\mathrm{d}z_{h}\mathrm{d}\boldsymbol{q}\,\mathrm{d}\tau}&=\sum_{q} \sigma_{0,q}(Q)   \int^{\gamma+i\infty}_{\gamma-i\infty} \frac{\mathrm{d}u}{2\pi i}\, e^{u\tau}  H(Q^2,\mu^2)\, J\Bigl(\frac{u}{Q^{2}},\mu\Bigr)\,  S_{\mathrm{thr}}\Bigl(\frac{u}{Q},\mu\Bigr)\, 
\mathcal{G}_{q\rightarrow h}\Bigl(\frac{u}{Q^{2}},z_{h},\boldsymbol{q},\mu\Bigr)
\,,\end{align}
where the sum on $q$ is over quark flavors. The variable $u$ is the Laplace conjugate of $\tau$, turning the convolutions in the factorization theorem into multiplications. The various functions entering this factorization are: The hard function $H$, encoding the hard scattering process $e^+ e^- \to q \bar q$ plus virtual corrections. The fragmenting jet function $\mathcal{G}$, describing the fragmenting hadron and corresponding contribution to the thrust measurement. The jet function $J$, accounting for the collinear contribution to  thrust measurement from the other hemisphere. The soft function $S_{\rm thr}$, describing the contribution to thrust from soft radiation. 
Explicit  one-loop expressions for $H$ and $J$ are given by~\cite{Manohar:2003vb,Bauer:2003di,Bauer:2003pi} 
\begin{align}
    H(Q^{2},\mu^{2}_{H})&=1+a_{S}\biggl[ -2C_{F}\ln^{2}\Bigl(\frac{\mu^{2}_{H}}{Q^{2}}\Bigr)-6C_{F}\ln\Bigl(\frac{\mu^{2}_{H}}{Q^{2}}\Bigr)+C_{F}\Bigl(-16+\frac{7\pi^{2}}{3}\Bigr)  \biggr]\,,
\nn \\
J\Bigl(\frac{u}{Q^{2}}, \mu_{J}\Bigr)&=1+a_{S}\biggl[ 2C_{F}\ln^{2}\Bigl(\frac{ue^{\gamma_{E}}\mu_{J}^{2}}{Q^{2}}\Bigr)+3C_{F}\ln\Bigl(\frac{ue^{\gamma_{E}}\mu_{J}^{2}}{Q^{2}}\Bigr)+C_{F}\Bigl( 7-\frac{2\pi^{2}}{3} \Bigr) \biggr]\,.
\label{eq:HJ}
\end{align}

The cross section for region 2 ($\tau Q \ll q_{T} \ll \sqrt\tau Q$) is given by
\begin{align}
\frac{\mathrm{d}\sigma_{2}}{\mathrm{d}z_{h}\mathrm{d}\boldsymbol{q}\,\mathrm{d}\tau}&=\sum_{q} \sigma_{0,q}(Q)  \int \frac{\mathrm{d}\boldsymbol{b}}{(2\pi)^{2}} \int^{\gamma+i\infty}_{\gamma-i\infty} \frac{\mathrm{d}u}{2\pi i}\, e^{i\boldsymbol{b}\cdot \boldsymbol{q}+u\tau}\, H(Q^2,\mu^2)\, J\Bigl(\frac{u}{Q^{2}},\mu\Bigr)\,  S_{\mathrm{thr}}\Bigl(\frac{u}{Q},\mu\Bigr) 
\nonumber\\ &\quad \times 
{\mathscr{S}}\Bigl(\boldsymbol{b},\frac{u}{Q},\mu, \frac{\zeta}{Q^2}\Bigr)\, D_{q\rightarrow h}(z_{h},\boldsymbol{b},\mu,\zeta)\,, 
\end{align}
where $\boldsymbol{b}$ is the Fourier conjugate of $\boldsymbol{q}$. Due to the hierarchy between $q_T$ and $\sqrt\tau Q$, the fragmenting jet function factorizes into a TMD fragmentation function $D$ and a collinear-soft function $\mathscr{S}$, that contributes both to thrust and the transverse momentum measurement.

The cross section for region 1 ($\tau Q \sim q_{T} \ll \sqrt\tau Q$) is given by 
\begin{align}
\frac{\mathrm{d}\sigma_{1}}{\mathrm{d}z_{h}\mathrm{d}\boldsymbol{q}\,\mathrm{d}\tau}&=\sum_{q} \sigma_{0,q}(Q)  \int \frac{\mathrm{d}\boldsymbol{b}}{(2\pi)^{2}} \int^{\gamma+i\infty}_{\gamma-i\infty} \frac{\mathrm{d}u}{2\pi i}\, e^{i\boldsymbol{b}\cdot \boldsymbol{q}+u\tau}\,  H(Q^2,\mu^2)\, J\Bigl(\frac{u}{Q^{2}},\mu\Bigr) 
\nonumber\\ &\quad \times 
{S}_{\mathrm{hemi}}\Bigl(\boldsymbol{b},\frac{u}{Q},\mu,\frac{\zeta}{Q^2}\Bigr)\, D_{q\rightarrow h}(z_{h},\boldsymbol{b},\mu,\zeta) \,.
\end{align}
Compared to region 2, there is no longer a hierarchy between $\tau Q$ and $q_T$, such that the thrust soft function and collinear-soft function combine into a new soft function, that we denote with $S_{\rm hemi}$. Thus $S_{\rm hemi}$ accounts for both the thrust measurement and the impact of soft radiation (in the hemisphere containing the fragmenting hadron) on the TMD measurement.

In this paper we will focus on region 2, because it allows the separate resummation of logarithms of $\tau$ and $q_T/Q$, and we expect this to be  relevant for phenomenology.

\subsection{Soft functions}
\label{sec:soft}

Here we collect the definition, one-loop calculation and results of all the (collinear-)soft functions that enter in the factorization theorems above. 
We also show how to obtain the subtracted soft functions for the $\delta$-regulator, and verify against earlier results obtained using the $\eta$-regulator.

\subsubsection{Thrust soft function}

The thrust soft function is defined in terms of soft Wilson lines (see \eqref{eq:S_SIA}) as 
\begin{equation}
    S_{\mathrm{thr}}(k^{+})=\frac{1}{N_{c}}\left\langle 0\left|\operatorname{Tr}\Bigl[\overline{\mathcal{T}}\Bigl(S_{n}^{\dagger}(0) S_{\bar{n}}(0)\Bigr)\, \delta\bigl(k^{+}-\mathcal{P}_{1}^{+}-\mathcal{P}_{2}^{-}\bigr)\, \mathcal{T}\Bigl(S^{\dagger}_{\bar{n}}(0) S_{n}(0)\Bigr)\Bigr]\right| 0\right\rangle
\end{equation}
at the bare level.
Here ($\overline{\mathcal{T}}$) $\mathcal{T}$ denotes (anti-)time ordering and the operator $\mathcal{P}_{1}$ ($\mathcal{P}_{2}$) gives the momentum of the soft radiation going into the hemisphere defined by $p_{i}^{+} < p_{i}^{-}$ ($p_{i}^{+} > p_{i}^{-}$).

At one-loop order and working in $u$-space, this leads to the following integrals 
\begin{align}\nn
    S^{[1]}_{\mathrm{thr}}\Bigl(\frac{u}{Q}\Bigr)&=  \int_{0}^{\infty} \mathrm{d} \tau\, e^{-u \tau}\int {\rm d} k^{+} Q \delta(k^{+}-Q \tau)\, S^{[1]}_{\mathrm{thr}}(k^{+}) \\
    &=Q \frac{ g^{2} C_{F} }{(2 \pi)^{3-2 \epsilon}} \biggl(\frac{e^{\gamma_{E}}\mu^{2}}{4 \pi}\biggr)^{\epsilon} \int\! \mathrm{d}^{d} \ell \int_{0}^{\infty}\! {\rm d} \tau e^{-u \tau}\,
    \frac{\delta[\ell^{+} \Theta(\ell^{-} \!-\! \ell^{+}) \!+\! \ell^{-} \Theta(\ell^{+} \!-\! \ell^{-}) \!-\! Q \tau]\, \Theta(\ell_{0})\, \delta(\ell^{2})}{\ell^{+}\ell^{-}} \nn \\ & \quad + \text{h.c.}
\end{align}
The $Q$ dependence in the soft function arises purely due to our choice of the dimensionless variable $u$, as highlighted by writing the soft function's argument as $u/Q$.
The result is 
\begin{equation}
S^{[1]}_{\mathrm{thr}}\Bigl(\frac{u}{Q}\Bigr)=-8a_{S}C_{F}\left(e^{\gamma_{E}}\mu^{2}\right)^{\epsilon} \frac{\Gamma(-\epsilon)\Gamma(-2\epsilon)}{\Gamma(1-\epsilon)^{2}}\Bigl(\frac{u}{Q}\Bigr)^{2\epsilon}\,, 
\end{equation}
where $a_s = \alpha_s/(4\pi)$.
The renormalized soft function is given by the remaining finite terms after expanding in $\epsilon$, 
\begin{equation}
    S_{\mathrm{thr}}\Bigl(\frac{u}{Q},\mu\Bigr)=1-a_{S}C_{F}\Big[\pi^{2}+8\ln^{2}\left( \frac{ue^{\gamma_{E}}\mu}{Q} \right)\Big]+\mathcal{O}(a_s^2)\,.
\end{equation}
The one-loop thrust soft function was calculated in refs.~\cite{Schwartz:2007ib,Fleming:2007xt}, and we agree with their result.

\subsubsection{Hemisphere soft function}

The new hemisphere soft function for this process has not been defined as a matrix element in the literature before, and only been extracted from consistency. At the bare level it is defined as
\begin{equation}
    S_{\mathrm{hemi}}(\boldsymbol{k},k^{+})=\frac{1}{N_{c}}\left\langle 0\left|\operatorname{Tr}\Bigl[\overline{\mathcal{T}}\Bigl(S_{n}^{\dagger}(0) S_{\bar{n}}(0)\Bigr)\, \delta\bigl(k^{+}-\mathcal{P}_{1}^{+}-\mathcal{P}_{2}^{-}\bigr)\, \delta^{2}(\boldsymbol{k}-\bm{\mathcal{P}}_{2}) \, \mathcal{T}\Bigl(S^{\dagger}_{\bar{n}}(0) S_{n}(0)\Bigr)\Bigr]\right| 0\right\rangle\,,
\end{equation}
where we have assumed that the final state hadron goes into the hemisphere 2.
The calculation is similar to that of the fully-unintegrated soft function from refs.~\cite{Procura:2014cba,Lustermans:2019plv}. The only difference is that we restrict the transverse momentum measurement to radiation going into one hemisphere, i.e.~for the emission of a single gluon with momentum $l$
\begin{align}
    &\delta^{2}(\boldsymbol{l}-\boldsymbol{k})=\delta^{2}(\boldsymbol{l}-\boldsymbol{k})\,[\Theta(l^{-}-l^{+})+\Theta(l^{+}-l^{-}) ]
    \nn \\ & \quad
    \longrightarrow \quad \delta^{2}(\boldsymbol{l}-\boldsymbol{k})\,\Theta(l^{+}-l^{-})+ \delta^{2}(\boldsymbol{k})\,\Theta(l^{-}-l^{+}) \,.
\end{align}
Dimensional regularization is not sufficient to regulate all divergences as this soft function suffers from light-cone singularities (rapidity divergences), which we regulate using the $\delta$-regulator~\cite{Chiu:2009yx,Chay:2012mh,Echevarria:2015byo,Echevarria:2016scs}. 
In the standard TMD case, where there are two collinear directions, one uses a separate $\delta^+$ and $\delta^-$ regulator to handle the rapidity divergences associated with the separation between the soft and each of these collinear directions. In our case there is only one collinear direction with rapidity divergences, so we will simply write $\delta$.
For the one-loop computation we only need to consider real radiation diagrams, since the virtual diagrams are proportional to terms of the form $\delta^\epsilon$ and can therefore be neglected in the limit $\delta\to 0$. This leads to 
\begin{align}
S^{[1]}_{\mathrm{hemi}}\Bigl(\boldsymbol{b}, \frac{u}{Q} \Bigr)&= \int \mathrm{d} \boldsymbol{k}\; e^{-\mathrm{i} \boldsymbol{b} \cdot \boldsymbol{k} }\int_{0}^{\infty} {\rm d} \tau\, e^{-u \tau}\int {\rm d} k^{+}\, Q\delta\bigl(k^{+}-Q \tau\bigr) S^{[1]}_{\mathrm{hemi}}\bigl(k^{+}, \boldsymbol{k}  \bigr) 
\nn \\
&=\frac{ g^{2} C_{F} }{(2 \pi)^{3-2 \epsilon}} \biggl(\frac{e^{\gamma_{E}}\mu^{2}}{4 \pi}\biggr)^{\epsilon}\int \mathrm{d}^{d} \ell \int_{0}^{\infty} {\rm d} \tau\, e^{-u \tau} 
\frac{\Theta(\ell_{0})\, \delta(\ell^{2})}{(\ell^{+}+\mathrm{i} Q \delta)(\ell^{-}-\mathrm{i} Q \delta)} 
\nonumber\\
& \quad 
\times
\bigl[
e^{-\mathrm{i} \boldsymbol{b} \cdot \boldsymbol{\ell}}\, 
\delta(\ell^{-}-Q \tau)\, Q\Theta(\ell^{+}-\ell^{-}) 
+ \delta(\ell^{+} -Q \tau)\, Q\Theta(\ell^{-}-\ell^{+}) \bigr]+\text{h.c.}
\end{align}
After performing the integrals, the result is 
\begin{align} 
    S^{[1]}_{\mathrm{hemi}}\Bigl(\boldsymbol{b},\frac{u}{Q} \Bigr)&=-4a_{S}C_{F}\left(e^{\gamma_{E}}\mu^{2}\right)^{\epsilon}\Gamma(-\epsilon)\biggl[ \ln (ue^{\gamma_{E}}\delta)B^{2\epsilon} 
    \nonumber\\ & \quad 
    + \frac{u^{2\epsilon}\Gamma(-2\epsilon)}{Q^{2\epsilon}\Gamma(1-\epsilon)^{2}}\Big(1+{}_{3}{F}_{2}\biggl(-\epsilon,-\epsilon,\frac{1}{2}-\epsilon;1-\epsilon,1-\epsilon ;-\frac{4B^{2} Q^{2}}{u^2}\biggr) \Big)\biggr]
\,,\end{align}
where $B^{2}=\boldsymbol{b}^{2}/4$ and $_{3} {F}_{2}$ is a hypergeometric function.
The remaining finite terms after expanding in $\epsilon$ are
\begin{align} \label{eq:S_1_finite}
    S^{[1]}_{\mathrm{hemi}}\Bigl(\boldsymbol{b},\frac{u}{Q},\mu,\delta^{2} Q^2\Bigr)&=4a_{S}C_{F}\biggl[ \frac{1}{2}\Big(-\ln(\mu^{2}B^{2}e^{2\gamma_{E}})\ln(\frac{\mu^{2}}{Q^{2}\delta^{2}})+ \frac{1}{2}\ln^{2}(\mu^{2}B^{2}e^{2\gamma_{E}})+\frac{\pi^{2}}{12}\Big)
    \nonumber\\
    &\quad
    +\frac{B^{2}Q^{2}F}{u^{2}}-\ln^{2}\Big(\frac{u}{BQ}\Big)-\ln^{2}\Big(\frac{ u e^{\gamma_{E}}\mu}{Q}\Big)-\frac{7\pi^{2}}{24} \biggr] \,,
\end{align}
where 
\begin{equation} \label{eq:F}
F={}_{4}F_{3}\biggl(\frac{3}{2}, 1, 1, 1; 2, 2, 2; -\frac{4 B^2 Q^2}{u^2}\biggr)
\,,\end{equation}
can be expressed in terms of harmonic polylogarithms.

\subsubsection{Thrust-TMD collinear-soft function}

Finally we turn to the thrust-TMD collinear-soft function, which at the bare level is defined as the following matrix element 
\begin{align}
    \mathscr{S}(\boldsymbol{k},k^{+}) = \frac{1}{N_{c}}\left\langle 0\left|\operatorname{Tr}\Bigl[\overline{\mathcal{T}}\Bigl(X_{n}^{\dagger}(0) V_{{n}}(0)\Bigr)\, \delta\bigl(k^{+}-\mathcal{P}^+\bigr)\, \delta^{2}(\boldsymbol{k}-\bm{\mathcal{P}}) \, \mathcal{T}\Bigl(V^{\dagger}_{n}(0) X_{n}(0)\Bigr)\Bigr]\right| 0\right\rangle\,,    
\end{align}
in terms of collinear-soft Wilson lines $X_n$ and $V_n$ defined in eq.~\eqref{eq:VX}. Because collinear-soft radiation is only emitted into the hemisphere in which the fragmenting hadron is measured this simplifies the thrust and transverse momentum contribution, i.e.~we do not need $\mathcal{P}_{1,2}$ and can simply write $\mathcal{P}$. 

At one-loop order the collinear-soft function is given by
\begin{align}
\mathscr{S}^{[1]}\Bigl(\boldsymbol{b}, \frac{u}{Q}\Bigr)&= \int \mathrm{d} \boldsymbol{k}\, e^{-\mathrm{i} \boldsymbol{b} \cdot \boldsymbol{k}} \int_{0}^{\infty} \mathrm{d} \tau\, e^{-u \tau} \int {\rm d} k^{+}\, Q \delta(k^{+}-Q \tau)\, \mathscr{S}^{[1]}(k^{+}, \boldsymbol{k})  \\
&
=\! \frac{ g^{2} C_{F} }{(2 \pi)^{3-2 \varepsilon}} \biggl(\frac{e^{\gamma_{E}}\mu^{2}}{4 \pi}\biggr)^{\epsilon} \!\int\! \mathrm{d}^{d} \ell \int_{0}^{\infty}\! {\rm d} \tau\, e^{-u \tau} e^{-\mathrm{i}  \boldsymbol{b} \cdot \boldsymbol{\ell} }\, Q\delta(\ell^{+}-Q \tau)  \frac{\Theta(\ell_{0})\,\delta(\ell^{2})}{\left(\ell^{+}\!+\!\mathrm{i} Q \delta\right)\left(\ell^{-}\!-\!\mathrm{i} Q \delta\right)} \!+\! \text{h.c.},
\nonumber
\end{align}
which results in
\begin{align}\nn
    \mathscr{S}^{[1]}\Bigl(\boldsymbol{b}, \frac{u}{Q} \Bigr)&=-4a_{S}C_{F}\left(e^{\gamma_{E}}\mu^{2}B^2\right)^{\epsilon}\Gamma(-\epsilon) \ln (ue^{\gamma_{E}}\delta)\,,\\
    &=4a_{S}C_{F}\ln (ue^{\gamma_{E}}\delta)
    \Bigl[\frac{1}{\epsilon}+\ln(\mu^2B^2)\Bigr]+\mathcal{O}(\epsilon^2)\,.
\end{align}
This result was first obtained  in~\cite{Procura:2014cba} (though not in conjugate space).

\subsubsection{Soft subtraction}
\label{sec:softsubtraction}

The factorization theorems of ref.~\cite{Makris:2020ltr} are expressed in terms of the unsubtracted TMDFF. To address the overlap between the collinear and soft radiation, we still have to perform a zero-bin subtraction~\cite{Manohar:2006nz}. This zero-bin is equal to the TMD soft function, which for the $\delta$-regulator is up to one loop given by~\cite{Echevarria:2011epo,Echevarria:2015byo} 
\begin{equation}
    S(\boldsymbol{b},\mu,\delta^2\zeta)=1+4a_{S}C_{F}\Big[-\ln(\mu^{2}B^{2}e^{2\gamma_{E}})\ln(\frac{\mu^{2}}{\delta^2 \zeta})+\frac{1}{2}\ln^{2}(\mu^{2}B^{2}e^{2\gamma_{E}})+\frac{\pi^{2}}{12}\Big] + \mathcal{O}(a_s^2)
\,.\end{equation}
To match the standard definition of the TMDFF, the square root of the zero-bin is absorbed in it 
\begin{equation}
    D_{q\rightarrow h}\Bigl(z_{h},\boldsymbol{b},\mu,\frac{\zeta}{Q^2}\Bigr)
    =\frac{D_{q\rightarrow h}(z_{h},\boldsymbol{b},\mu,\delta^{2}Q^2)}{\sqrt{S_{}(\boldsymbol{b},\mu,\delta^{2}\zeta)}}\,,
\end{equation}
removing the rapidity divergences encoded by $\delta$. The other square root of the zero-bin is absorbed into the definition of the other rapidity-divergent soft matrix element present in the factorization.

For region 1, this other object is the hemisphere soft function. Including a square root of the zero-bin cancels the $\delta$'s, resulting in the following expression for the renormalized function up to NLO, 
\begin{align}
    S_{\mathrm{hemi}}\Bigl(\boldsymbol{b}, \frac{u}{Q},\mu,\frac{\zeta}{Q^2}\Bigr)&= \frac{S_{\mathrm{hemi}}(\boldsymbol{b}, u/Q,\mu,\delta^{2}Q^2)}{\sqrt{S_{}(\boldsymbol{b},\mu,\delta^{2}\zeta)}}
    \\ &
  \hspace{-3cm}
    =1+4a_{S}C_{F} \biggl[ -\frac{1}{2}\ln(\mu^{2}B^{2}e^{2\gamma_{E}})\ln\Bigl(\frac{\zeta}{Q^{2}}\Bigr) +\frac{B^{2}Q^{2}F}{u^{2}}-\ln^{2}\Bigl(\frac{u}{BQ}\Bigr)-\ln^{2}\Bigl(\frac{ u e^{\gamma_{E}}\mu}{Q}\Bigr)-\frac{7\pi^{2}}{24} \biggr]  + \mathcal{O}(a_s^2)\,, 
    \nonumber
\end{align}
with $F$ given in eq.~\eqref{eq:F}. Consistency of the anomalous dimensions requires $\zeta=Q^2$.\footnote{With two collinear directions, the corresponding equation is $\zeta \bar \zeta = Q^4$.}
For region 2, the collinear-soft function plays this role instead. Its renormalized form at one-loop order is 
\begin{align}
    {\mathscr{S}}\Bigl(\boldsymbol{b}, \frac{u}{Q},\mu,\frac{\zeta}{Q^2}\Bigr) &= \frac{\mathscr{S}(\boldsymbol{b},u/Q,\mu,\delta^{2}Q^2)}{\sqrt{S_{}(\boldsymbol{b},\mu,\delta^{2}\zeta)}} \\
    &=1+2a_{S}C_{F}\Big[2\ln(\frac{u e^{\gamma_{E}}\mu}{\sqrt\zeta}) \ln(\mu^{2}B^{2}e^{2\gamma_{E}})-\frac{1}{2}\ln^{2}(\mu^{2}B^{2}e^{2\gamma_{E}})-\frac{\pi^{2}}{12}\Big] + \mathcal{O}(a_s^2)
\,.\nonumber\end{align}
We have compared these results to \cite{Makris:2020ltr}, finding agreement. There the $\eta$-regulator was used to regulate rapidity divergences, for which the zero-bin vanishes.

\section{Factorization for SIDIS}
\label{sec:SIDIS}

Next we will discuss the corresponding measurement in semi-inclusive deep inelastic scattering (SIDIS).
Where in $e^+e^-$ we considered the transverse momentum of a final-state hadron with respect to the thrust axis, in this case we consider the transverse momentum of the \emph{initial} hadron with respect to the 1-jettiness (or thrust) axis denoted by $q_J$.\footnote{In SIDIS one usually measures a final-state hadron, while here we measure an axis instead.}

We will now discuss the kinematics of this process, which we write as
\begin{align}
h(P) + \ell (l) \longrightarrow \ell(l') + J(p_J, q_J)+X\,.
\end{align}
indicating the momenta of the particles, the jet momentum $p_J$ and the axis $q_J$, which are not necessarily aligned.
The momentum of the virtual photon is $q=l-l'$, 
\begin{align}
Q^2=-q^2 > 0\,.
\end{align}
and the other relevant kinematic variables are
\begin{align}
\qquad
x=\frac{Q^2}{2P \cdot q},\qquad y=\frac{P \cdot q}{P \cdot l}\,,
\end{align}
with all masses neglected.
Traditionally, the kinematics of SIDIS is defined with respect to the system of vectors $P$ and $q$~\cite{Bacchetta:2006tn}. The plane perpendicular ($\perp$) to the hadron momentum $P$ and the virtual photon $q$ is defined via the tensor
\begin{align}
\label{eq:def_perp}
g_\perp^{\mu\nu}&=g^{\mu\nu}-\frac{2 x}{Q^2}(2 x P^\mu P^\nu+q^\mu P^\nu+P^\mu q^\nu)\,,
\qquad g_\perp^{\mu\nu}g_{\perp,\mu\nu}=2\,, \nn \\
\epsilon^{\mu\nu}_\perp&=
\frac{2 x}{Q^2}\epsilon^{\mu\nu\alpha\beta }P_\alpha q_\beta\,,\qquad 
\epsilon^{\mu\alpha}_\perp
\epsilon^\nu_{\perp,\alpha}=g^{\perp,\mu\nu}\,, \text{ with }\epsilon^{0123}=-\epsilon_{0123}=1\,.
\end{align}
We will also use the plane transverse (T) to $P$ and the 1-jettiness axis $q_J$, defined in a similar way
\begin{align} \label{eq:def_T}
    g^{\mu\nu}_T&=g^{\mu\nu}-\frac{1}{Pq_J}\Bigl(\frac{-q_J^2}{Pq_J}P^\mu P^\nu + q_J^\mu P^\nu + P^\mu q_J^\nu\Bigr)\,, \qquad g_T^{\mu\nu}g_{T,\mu\nu}=2\,, \nn \\
\epsilon^{\mu\nu}_T&=
\frac{1}{Pq_J}\epsilon^{\mu\nu\alpha\beta }P_\alpha q_{J,\beta}\,,\qquad 
\epsilon^{\mu\alpha}_T
\epsilon^\nu_{T,\alpha}=g^{T,\mu\nu}\,.
\end{align}

\subsection{1-jettiness}
In SIDIS one usually has a single jet in the final state, making it natural to describe it using 1-jettiness~\cite{Stewart:2010tn}
\begin{align}
\label{eq:tauX}
    \tau=\frac{1}{E_\tau^2}\sum_{i}\text{min}\{q_B\cdot p_i,\, q_J\cdot p_i\}\,.
\end{align}
Here $q_B$ is a reference vector for initial state or beam direction, $E_\tau$ is an energy scale used for  normalization, and $q_J$ characterizes the jet direction, which we will consider as the analogue of the thrust axis in $e^+e^-$. 
In the following we will use $E_\tau^2=Q^2/2$  and $q_B=xP$, with $x$ the initial parton momentum fraction. 

We will consider two different ways of defining $q_J$, see fig.~\ref{fig:SIDIScases}, and present the corresponding factorization theorems. 
This follows the factorization analysis of ref.~\cite{Kang:2013nha} where the transverse momentum is not measured.
In case \textit{a} (the aligned case), $q_J$ can be obtained either by making the jet momentum $p_J$ obtained through a jet algorithm massless,
$q_J^a=p_J^0(1,\vec{p}_J/|\vec{p}_J|)$, or through the minimization of the direction $\hat n_J^a$ in $q_J^a=|\vec q_J^{\,a}|(1,\hat n_J^a)$, and
\begin{align}
\label{eq:tauA}
    \tau^a=\frac{2}{Q^2}\underset{\hat n_J^a}{\text{min}}\sum_{i\in X}\text{min}\{xP \cdot p_i,\, q_J^a\cdot p_i\}\,.
\end{align}
Both choices have the property that $q_J^a = p_J + \mathcal{O}(\lambda^2)$, i.e.~$q_J^a$ and $p_J$ are aligned.

In case \textit{b} we consider specifically the hadron Breit frame where the virtual photon is spacelike, and choose $q_J^b = q+xP$ such that $q_J^b = p_J + \mathcal{O}(\lambda)$. In this case
\begin{align}
\label{eq:tauB}   
    \tau^b=\frac{2}{Q^2}\sum_{i\in X}\text{min}\{xP \cdot p_i,\, q_J^b\cdot p_i\}\,.
\end{align}

In the following we show that  the vector boson transverse momentum $\boldsymbol{q}_T$ is the relevant observable for the factorization analysis of case \textit{a}.
For case \textit{b} we have  $g^{\mu\nu}_T=g^{\mu\nu}_\perp$
and the factorization holds for the transverse component $p_{JT} = p_{J\perp}$ of the jet momentum
\begin{equation} \label{eq:PJ}
p_J^\mu = \sum_i \Theta(q_J\cdot p_i<q_B \cdot p_i)\, p_i^\mu
\,,
\end{equation}
using the partitioning of the phase space provided by eq.~\eqref{eq:tauX}.

\subsection{Factorization for  case \textit{a}}

In case \textit{a} the jet momentum $p_J$ and the axis specified by $q_J^a$ are almost aligned.  
We have\footnote{Technically this is label momentum conservation in SCET, which explains why there is only $r_{JT}$.}
\begin{align}
 q_J^{a\,\mu}=q^\mu+xP^\mu+r_{JT}^\mu\,,
\end{align}
where a nonzero $r_{JT}^\mu=-q_T^\mu$ of $\mathcal{O}(Q\lambda)$ arises due to transverse momenta of quarks in the initial hadron. There is no transverse momentum contribution associated with the jet because $T$ means transverse to $q_J^a$, see eq.~\eqref{eq:def_T}, so $q_{JT}^a=0$.

For region 3 (that is $\tau Q \ll q_T \sim \sqrt{\tau} Q$) we then have 
\begin{align}
\label{eq:Xr3}
\frac{\mathrm{d}\sigma_{3}}{\mathrm{d}x\,\mathrm{d}Q^{2}\,\mathrm{d}\boldsymbol{q}\,\mathrm{d}\tau}&=\sum_{q} \sigma_{0,q}(y,Q)   \int^{\gamma+i\infty}_{\gamma-i\infty} \frac{\mathrm{d}u}{2\pi i}\, e^{u\tau}  H^\text{DIS}(Q^2,\mu^2)\, J\Bigl(\frac{u}{Q^{2}},\mu\Bigr)\,  S_{\mathrm{thr}}\Bigl(\frac{u}{Q},\mu\Bigr)
\nonumber\\ & \quad \times 
\mathcal{B}_{q\leftarrow h} \Bigl(\frac{u}{Q^{2}},x,\boldsymbol{q},\mu\Bigr) \,,
\end{align}
where the sum runs over all quark flavors $q=u, \bar u, d,\dots$. Compared to eq.~\eqref{eq:Xsec3i}  the hard function is changed, $H^\text{DIS}(Q^2,\mu^2)=\text{Re}(H(-Q^2,\mu^2))$, and the fragmenting jet function $\mathcal{G}$ is replaced by the corresponding beam function $\mathcal{B}_{q\leftarrow h}$. This is the Laplace transform of the beam function double differential in thrust and transverse momentum~\cite{Jain:2011iu,Gaunt:2014xxa}.
 
Extending to region 2 ($\tau Q\ll q_T\ll \sqrt{\tau} Q $), the cross section  is given by 
\begin{align}
\label{eq:Xr2}
\frac{\mathrm{d}\sigma_{2}}{\mathrm{d}x\,\mathrm{d}Q^{2}\,\mathrm{d}\boldsymbol{q}\,\mathrm{d}\tau}&=\sum_{q} \sigma_{0,q}(y,Q)  \int \frac{\mathrm{d}\boldsymbol{b}}{(2\pi)^{2}} \int^{\gamma+i\infty}_{\gamma-i\infty} \frac{\mathrm{d}u}{2\pi i}\, e^{i\boldsymbol{b}\cdot \boldsymbol{q}+u\tau}\,  H^\text{DIS}(Q^2,\mu^2)\, J\Bigl(\frac{u}{Q^2},\mu\Bigr)   \nonumber\\
&\quad \times S_{\mathrm{thr}}\Bigl(\frac{u}{Q},\mu\Bigr)\,\mathscr{S}\Bigl(\boldsymbol{b},\frac{u}{Q},\mu, \frac{\zeta}{Q^2}\Bigr) F_{q\leftarrow h}(x,\boldsymbol{b},\mu,\zeta) \,.
\end{align}
The double differential beam function $\mathcal{B}_{q\leftarrow h}$ has been factorized into a collinear-soft function $\mathscr{S}$ and the TMDPDF $F_{q\leftarrow h}$,
in direct correspondence with the $e^+e^-$ case.

In region 1 ($\tau Q\sim q_T\ll \sqrt{\tau} Q$) the factorization reads 
\begin{align}
\label{eq:Xr1}
\frac{\mathrm{d}\sigma_{1}}{\mathrm{d}x\,\mathrm{d}Q^{2}\,\mathrm{d}\boldsymbol{q}\,\mathrm{d}\tau}&=\sum_{q} \sigma_{0,q}(y,Q)  \int \frac{\mathrm{d}\boldsymbol{b}}{(2\pi)^{2}} \int^{\gamma+i\infty}_{\gamma-i\infty} \frac{\mathrm{d}u}{2\pi i}\, e^{i\boldsymbol{b}\cdot \boldsymbol{q}+u\tau}\,  H^\text{DIS}(Q^2,\mu^2)\, J\Bigl(\frac{u}{Q^{2}},\mu\Bigr) \nonumber\\
& \quad \times S_{\mathrm{hemi}}\Bigl(\boldsymbol{b},\frac{u}{Q},\mu,\frac{\zeta}{Q^2}\Bigr) F_{q\leftarrow h}(x,\boldsymbol{b},\mu,\zeta) \,.
\end{align}
The matching of $S_{\mathrm{hemi}}$ in region 1 onto the thrust soft function $S_{\rm thr}$ and collinear-soft function $\mathscr{S}$ in region 2 is the same for SIDIS and $e^+e^-$.

All (collinear-) soft functions are defined similarly to the $e^+e^-$ case, with the only difference stemming from the direction of the Wilson lines, which are aligned with either the initial state or final state. This difference does not matter for the unpolarized case studied here, but there can be subtleties for polarized scattering~\cite{Echevarria:2012js,Echevarria:2014rua,Collins:2014jpa} that we leave for future work. The operator definition of the TMDPDF is well known, and higher-order perturbative contributions have been calculated, particularly for the unpolarized case~\cite{Catani:2013tia,Gehrmann:2014yya,Echevarria:2016scs,Luo:2019hmp,Duhr:2020seh,Ebert:2020yqt}.

\subsection{Factorization for  case \textit{b}
}

Next we consider the case where the axis is fixed through $q_J^b=q+xP$.
A direct consequence of this choice is that $q_{T}=0$, since $T$ means transverse to $q_J^b$ and $P$.
Instead we will consider $p_{JT} = p_{J\perp}$, since $g_{T}^{\mu\nu}=g_{\perp}^{\mu\nu}$ because of the definition of $q_J^b$, see eqs.~\eqref{eq:def_perp} and \eqref{eq:def_T}.
Interestingly, the cross section 
\begin{align}
     \frac{\text{d}\sigma}{\text{d}x\,\text{d}Q^2\,\text{d} \mathbf{p}_{JT}\,\text{d}\tau^b}\,,
\end{align} 
has the same formulae as for case \textit{a} with $\mathbf{q}_T\leftrightarrow \mathbf{p}_{JT}$, except for region 3.
For region 3 we have 
\begin{align}
\label{eq:Xr3b}
\frac{\mathrm{d}\sigma_{3}}{\mathrm{d}x\,\mathrm{d}Q^{2}\,\mathrm{d}\boldsymbol{p}_{JT}\,\mathrm{d}\tau}&=\sum_{q} \sigma_{0,q}(y,Q)  
\int^{\gamma+i\infty}_{\gamma-i\infty} \frac{\mathrm{d}u}{2\pi i}\, e^{u(\tau -p_{JT}^2/Q^2})  H^\text{DIS}(Q^2,\mu^2)\, J\Bigl(\frac{u}{Q^{2}},\mu\Bigr)\,  S_{\mathrm{thr}}\Bigl(\frac{u}{Q},\mu\Bigr)
\nonumber\\ & \quad \times 
\mathcal{B}_{q\leftarrow h} \Bigl(\frac{u}{Q^{2}},x,\boldsymbol{p}_{JT},\mu\Bigr) \,,
\end{align}
where the $\exp(-up_{JT}^2/Q^2)$ arises because the offset between the jet momentum $p_J$ and the axis $q_J$ directly contributes to the 1-jettiness measurement.

This involves the same double differential beam function as in case \textit{a}, though the transverse momentum is with respect to a different axis. In region 2 this offset is expanded because $p_{JT}^2\ll \tau Q^2$,
\begin{align}
\label{eq:Xr2b}
\frac{\mathrm{d}\sigma_{2}}{\mathrm{d}x\,\mathrm{d}Q^{2}\,\mathrm{d}\boldsymbol{p}_{JT}\,\mathrm{d}\tau}&=\sum_{q} \sigma_{0,q}(y,Q) 
\int \frac{\mathrm{d}\boldsymbol{b}}{(2\pi)^{2}} \int^{\gamma+i\infty}_{\gamma-i\infty} \frac{\mathrm{d}u}{2\pi i}\, e^{i\boldsymbol{b}\cdot \boldsymbol{p}_{JT}+u\tau}\,
H^\text{DIS}(Q^2,\mu^2)\, J\Bigl(\frac{u}{Q^2},\mu\Bigr)   \nonumber\\
&\times S_{\mathrm{thr}}\Bigl(\frac{u}{Q},\mu\Bigr)\,\mathscr{S}\Bigl(\boldsymbol{b},\frac{u}{Q},\mu, \frac{\zeta}{Q^2}\Bigr) F_{q\leftarrow h}(x,\boldsymbol{b},\mu,\zeta) \,.
\end{align}
and the beam function factorizes into a collinear-soft function and a TMDPDF. In region 1, we have 
\begin{align}
\label{eq:Xr1b}
\frac{\mathrm{d}\sigma_{1}}{\mathrm{d}x\,\mathrm{d}Q^{2}\,\mathrm{d}\boldsymbol{p}_{JT}\,\mathrm{d}\tau}&=\sum_{q} \sigma_{0,q}(y,Q)  
\int \frac{\mathrm{d}\boldsymbol{b}}{(2\pi)^{2}} \int^{\gamma+i\infty}_{\gamma-i\infty} \frac{\mathrm{d}u}{2\pi i}\, e^{i\boldsymbol{b}\cdot \boldsymbol{p}_{JT}+u\tau}\,
H^\text{DIS}(Q^2,\mu^2)\, J\Bigl(\frac{u}{Q^{2}},\mu\Bigr) \nonumber\\
& \quad \times S_{\mathrm{hemi}}\Bigl(\boldsymbol{b},\frac{u}{Q},\mu,\frac{\zeta}{Q^2}\Bigr) F_{q\leftarrow h}(x,\boldsymbol{b},\mu,\zeta) \,.
\end{align}
This region is special because it contains a TMDPDF whose transverse momentum can approach $\Lambda_{\rm QCD}$.

\section{Resummation}
\label{sec:resummation}

The factorization for $e^+e^-$ in sec.~\ref{sec:SIA} and SIDIS in sec.~\ref{sec:SIDIS} serves two purposes: to isolate universal nonperturbative ingredients such as TMDPDFs from perturbatively calculable ingredients, and to enable the resummation of large logarithms of $\tau$ and $q_{T}/Q$ by separating physics at different scales. We will accomplish this resummation by evaluating all ingredients at their natural scales and evolving them to a common scale. 

There are subtleties related to the choice of natural scales, for which we could either use the original variables $q_T$ and $\tau$, or the conjugate variables $b$ and $u$. Choosing scales in terms of the conjugate variables clearly removes all large logarithms due to the multiplicative structure of the RGE. While for transverse momentum resummation this is important as resummation in momentum space is challenging~\cite{Frixione:1998dw,Ebert:2016gcn}, 
for thrust also resummation in terms of $\tau$ has been considered.

Here we consider three approaches:
In the first we choose scales in terms of the conjugate variables $\boldsymbol{b}$ and $u$. While this clearly resums all logarithms, a drawback is that $\mu$ scales will depend on $u$, requiring a numerical inverse Laplace transform. In the second approach, we still use $\boldsymbol{b}$ but employ $\tau$ instead of $u$ in our choice of scales. 
A third possibility consists of choosing a common initial condition for all functions that have $u$-dependence, called the $\mathcal{M}$-scheme~\cite{Echevarria:2026vca}. This effectively turns off the resummation of logarithms of $\tau$, but benefits from having the same and simpler resummation structure as in the TMD case.

In sec.~\ref{sec:RGE} we will collect all ingredients of the renormalization group equations. The solutions of the RGE for the three schemes, together with the expressions for the resulting cross section, are reported in secs.~\ref{sec:indyscheme}
and \ref{sec:Mscheme}.

\subsection{Renormalization group equations}
\label{sec:RGE}

All anomalous dimensions are known to order $\alpha_s^3$, except for our $S_{\rm hemi}$ and the collinear-soft function $\mathscr{S}$, but they can be extracted from consistency relations.

The renormalization group equations for the hard, jet and thrust soft function are given by
\begin{align}
     \frac{\mathrm{d}}{\mathrm{d}\ln \mu}\,H(Q^2,\mu^2)&=\biggl[ -2\Gamma_{\mathrm{cusp}}(a_{s})\ln(\frac{\mu^2}{Q^2})+\gamma^{H}(a_{s}) \biggr] H(Q^2,\mu^2),
\nn \\
 \frac{\mathrm{d}}{\mathrm{d}\ln \mu}\,J\Bigl(\frac{u}{Q^{2}},\mu\Bigr)&=\biggl[ 2\Gamma_{\mathrm{cusp}}(a_{s})\ln(\frac{ue^{\gamma_{E}}\mu^2}{Q^2})+\gamma^{J}(a_{s}) \biggr] J\Bigl(\frac{u}{Q^{2}},\mu\Bigr),
\nn \\
     \frac{\mathrm{d}}{\mathrm{d}\ln \mu}\,S_{\mathrm{thr}}\Bigl(\frac{u}{Q},\mu\Bigr)&=\biggl[ -4\Gamma_{\mathrm{cusp}}(a_{s})\ln(\frac{ue^{\gamma_{E}}\mu}{Q})+\gamma^{S_{\mathrm{thr}}}(a_{s}) \biggr]\, S_{\mathrm{thr}}\Bigl(\frac{u}{Q},\mu\Bigr)\,.
\end{align}
We then have three functions with double-evolution equations, namely the 
TMDPDF and the TMD fragmentation function ($\Phi=F_{q\leftarrow h}, D_{q\rightarrow h}$),
\begin{align}
     \frac{\mathrm{d}}{\mathrm{d}\ln \mu}\,\Phi(x,\boldsymbol{b},\mu,\zeta)&=\gamma^{F}(\mu,\zeta)\, \Phi(x,\boldsymbol{b},\mu,\zeta), \nn \\
    \zeta\frac{\mathrm{d}}{\mathrm{d}\zeta}\,\Phi(x,\boldsymbol{b},\mu,\zeta)&=-\mathcal{D}(\boldsymbol{b},\mu)\,\Phi(x,\boldsymbol{b},\mu,\zeta)\,,
\end{align}
and the 
collinear-soft function, 
\begin{align}
     \frac{\mathrm{d}}{\mathrm{d}\ln \mu}\,\mathscr{S}\Bigl(\boldsymbol{b},\frac{u}{Q},\mu, \zeta\Bigr) &=\gamma^{\mathscr{S}}(u,\mu,\zeta)\, \mathscr{S}\Bigl(\boldsymbol{b},\frac{u}{Q},\mu, \zeta\Bigr), \nn \\
    \zeta\frac{\mathrm{d}}{\mathrm{d}\zeta}\,\mathscr{S}\Bigl(\boldsymbol{b},\frac{u}{Q},\mu, \zeta\Bigr)&=-\mathcal{D}(\boldsymbol{b},\mu)\, \mathscr{S}\Bigl(\boldsymbol{b},\frac{u}{Q},\mu, \zeta\Bigr)\,.
\end{align}
The Collins-Soper kernel $\mathcal{D}$ is discussed explicitly in sec.~\ref{sec:Dkernel}.
The  other anomalous dimensions  can be written as 
\begin{align}
\label{eq:gammaTMD}
\gamma^F(\mu,\zeta)&=
\Gamma_{\mathrm{cusp}}(a_{s})\ln(\frac{\mu^2}{\zeta})-\gamma^{V}(a_{s}) \,, \\
\gamma^{\mathscr{S}}(u,\mu,\zeta)&=
\Gamma_{\mathrm{cusp}}(a_{s})\ln\Bigl(\frac{u^2 e^{2\gamma_E} Q^2\mu^2}{\zeta}\Bigr)-\gamma^{\mathscr{S}}(a_{s}) \,.
\label{eq:gammaCS}
\end{align}
The expressions for the anomalous dimensions and the QCD $\beta$-function, in the $\overline{{\rm MS}}$ renormalization scheme can be expanded as 
\begin{align}
\Gamma_{\rm cusp} &=
\sum_{n=1}^{\infty} \Gamma_{n-1}  a_s^n
\,,
\qquad
\gamma^i = \sum_{n=1}^{\infty} \gamma^i_{n-1} a_s^n
\,,
\qquad
\beta= -2\alpha_s \sum_{n=1}^\infty \beta_{n-1} a_s^n
\,,
\end{align}
whose coefficients are collected in appendix~\ref{appendix:ad}.

In this work the resummation is carried out at next-to-next-to-leading logarithmic (NNLL) order, which includes the cusp anomalous dimension up to three loop order, and the other anomalous dimensions at two loops.

The resummed hard function is the same for all schemes that we present below, and is given by 
\begin{align}
     H(Q^2,\mu^2)&=H(Q^2,\mu_{H}^2)\exp\Bigl[ 4S(\mu_{H},\mu)-A_{H}(\mu_{H},\mu)  \Bigr] \Big(\frac{\mu_{H}^{2}}{Q^2}\Big)^{2A_{\Gamma}(\mu_{H},\mu)}.
\end{align}
The one-loop expression for the boundary condition $H(Q^2,\mu_{H})$ is given in eq.~(\ref{eq:HJ}).    
The resummation of the remaining ingredients in the factorization theorem is discussed in the following subsections.

\subsection{Evolution of functions}
\label{sec:indyscheme}

The functions $J$ and $S_{\rm thr}$ can be resummed in the  standard way, see e.g.~\cite{Becher:2006mr,Becher:2008cf}, 
\begin{align}
 J\Bigl(\frac{u}{Q^2},\mu\Bigr)&=J\Bigl(\frac{u}{Q^2},\mu_{J}\Bigr)\,
R_J\Bigl[\frac{u}{Q^2},\mu_J\rightarrow \mu\Bigr]\,,
\nn \\ 
S_{\mathrm{thr}}\Bigl(\frac{u}{Q},\mu\Bigr)&=S_{\mathrm{thr}}\Bigl(\frac{u}{Q},\mu_{S_{\mathrm{thr}}}\Bigr)\,
R_{S_\text{thr}}\Bigl[\frac{u}{Q},\mu_{S_\text{thr}}\rightarrow \mu\Bigr]\,.
\end{align}
The evolution kernels entering these expressions are given by
\begin{align}
R_J\Bigl[\frac{u}{Q^2},\mu_J\rightarrow \mu\Bigr]&=\exp\bigl[ -4S(\mu_{J},\mu)-A_{J}(\mu_{J},\mu)  \bigr] \Bigl(\frac{ue^{\gamma_E}\mu_{J}^{2}}{Q^2}\Bigr)^{-2A_{\Gamma}(\mu_{J},\mu)}, \nn \\
R_{S_\mathrm{thr}}\Bigl[\frac{u}{Q},\mu_{S_\text{thr}}\rightarrow \mu\Bigr]&=\exp\bigl[ 4S(\mu_{S_{\mathrm{thr}}},\mu)-A_{S}(\mu_{S_{\mathrm{thr}}},\mu) \bigr] \Bigl(\frac{u e^{\gamma_E}\mu_{S_{\mathrm{thr}}}}{Q}\Bigr)^{4A_{\Gamma}(\mu_{S_{\mathrm{thr}}},\mu)}\,,
\end{align}
with $S$ and $A$ integrals of the anomalous dimensions, whose explicit form we report in appendix~\ref{sec:appendixa}.

The initial renormalization scales are chosen to be the natural scale for each function, minimizing the logarithms in the fixed-order expression for each function. Here we will consider two different resummation schemes, depending on whether we choose the scales in $u$-space or $\tau$-space, with the $\mathcal{M}$-scheme deferred to sec.~\ref{sec:Mscheme}. Concretely,
\begin{align}
    \text{$u$-scheme:} \quad \mu_{J}&=\frac{Q}{\sqrt{ue^{\gamma_{E}}}}, \quad \mu_{S_{\mathrm{thr}}}=\frac{Q}{ue^{\gamma_{E}}}, \nn \\
    \text{$\tau$-scheme:} \quad
    \mu_{J}&=\sqrt{\tau}Q, \qquad \mu_{S_{\mathrm{thr}}}=\tau Q.
\end{align}
In the first case, the large logarithms are canceled before the inverse Laplace transform takes place, which helps with the convergence of the numerical implementation, whereas in the second case they are resummed after the inverse Laplace transformation.
Currently, \texttt{artemide} is not able to handle complex scales in $\alpha_s$, as needed for the inverse Laplace transform, see sec.~\ref{sec:laplace}.
We therefore modify the $u$-scheme such that the scale in $\alpha_s$ is always in terms of $\tau$.

For TMDs  we use the $\zeta$-prescription of ref.~\cite{Scimemi:2018xaf} that is implemented in the \texttt{artemide} code. Using a common notation $\Phi(x,\boldsymbol{b})$ for the unpolarized TMDPDF and TMDFF, and omitting flavor labels for brevity,
\begin{align}
\label{eq:RTMD}
    \Phi(x,\boldsymbol{b},Q,Q^2)&=
    R[\boldsymbol{b},(\mu_{0},\zeta_{0})\rightarrow (Q,Q^2)]\,\Phi(x,\boldsymbol{b})\,, \nn \\
    R[\boldsymbol{b},(\mu_{0},\zeta_{0})\rightarrow (Q, Q^2)]&= 
\left( \frac{Q^2}{\zeta_{\mu}(b) } \right)^{-\mathcal{D}(\boldsymbol{b}, Q)}.
\end{align}
The scales $(\mu,\zeta_\mu(b))$ belong to the special equi-potential line  of the evolution (see~\cite{Scimemi:2018xaf}) defined by,
\begin{equation}\label{def:zeta-line}
\Gamma_{\text{cusp}}(\mu)\,\ln\Bigl(\frac{\mu^2}{\zeta_\mu(b)}\Bigr)-\gamma^V(\mu)=2\mathcal{D}(\boldsymbol{b},\mu)\, \frac{{\rm d} \ln\zeta_\mu(b)}{{\rm d}\ln\mu^2}\,,
\end{equation}
with boundary condition that $(\mu_0,\zeta_0)$ is the saddle point, 
\begin{equation}\label{th:saddle-point}
\mathcal{D}(\boldsymbol{b},\mu_0)=0,\qquad \gamma^F(\mu_0,\zeta_0)=0.
\end{equation}
These conditions imply that the optimal TMDs are exactly scale-independent (for any $\mu$ and $b$) and can therefore be denoted without scales,
\begin{equation}
\Phi(x,\boldsymbol{b})\equiv \Phi(x,\boldsymbol{b},\mu,\zeta_\mu(b)).
\end{equation}
A similar reasoning to the TMD case also works for the collinear-soft function. Eq.~\eqref{eq:gammaTMD} and \eqref{eq:gammaCS} are equivalent with the replacements $(\zeta,\gamma^V)\leftrightarrow (\tilde\zeta,\gamma^\mathscr{S})$ and $\tilde\zeta=\zeta/(u^2 e^{2\gamma_E})$.
This leads to 
\begin{align}
   \mathscr{S}\Bigl(\boldsymbol{b},\frac{u}{Q},Q,\frac{Q^2}{u^2 e^{2\gamma_E}}\Bigr) &=R_{{\mathscr{S}}}\Bigl[\boldsymbol{b},\frac{u}{Q}, (\mu_0, \tilde{\zeta}_{\mathscr{S}})\rightarrow \Bigl(Q,\frac{Q^2}{u^2 e^{2\gamma_E}}\Bigr)\Bigr]\mathscr{S}\Bigl(\boldsymbol{b},\frac{u}{Q}\Bigr)\,, \nn \\
    R_{{\mathscr{S}}}\Bigl[\boldsymbol{b}, \frac{u}{Q},(\mu_0, \tilde{\zeta}_{\mathscr{S}})\rightarrow \Bigl(Q,\frac{Q^2}{u^2 e^{2\gamma_E}}\Bigr)\Bigr] &=\Bigl( \frac{u^2 e^{2\gamma_E} \tilde{\zeta}_{\mu}(\mu_0,\tilde{\zeta}_{{\mathscr{S}}})}{Q^2} \Bigr)^{+\mathcal{D}(\boldsymbol{b},Q)} \,,
\end{align}
where in direct analogy to the TMD case the function $\mathscr{S}(\boldsymbol{b},\frac{u}{Q})$ is scaleless and fixed at the saddle point.
Indeed, the equi-potential curve and saddle point are obtained in the same way as in the TMD case, 
\begin{align}
&\Gamma_{\text{cusp}}(\mu)\,\ln\Bigl(\frac{\mu^2}{\tilde\zeta_\mu(b)}\Bigr)-\gamma^{\mathscr{S}}(\mu)=2\mathcal{D}(\boldsymbol{b},\mu)\,\frac{{\rm d} \ln\tilde\zeta_\mu(b)}{{\rm d} \ln\mu^2}\,, \nn \\
&\qquad \mathcal{D}(\boldsymbol{b},\mu_0)=0,\qquad \gamma^\mathscr{S}(\mu_0,\tilde\zeta_\mathscr{S})=0\,.
\end{align}
Putting everything together, for region 2 the cross-section is given by 
\begin{align}
\frac{\mathrm{d}\sigma_{2}}{\mathrm{d}z_{h}\,\mathrm{d}\boldsymbol{q}\,\mathrm{d}\tau}&=\sum_{q} \sigma_{0,q}(Q)  \int \frac{\mathrm{d}\boldsymbol{b}}{(2\pi)^{2}} \int^{\gamma+i\infty}_{\gamma-i\infty} \frac{\mathrm{d}u}{2\pi i}\, e^{i\boldsymbol{b}\cdot \boldsymbol{q}+u\tau}  H(Q^2,\mu^2=Q^2)\,  
 J\Bigl(\frac{u}{Q^2},\mu_J\Bigr)\, S_{\mathrm{thr}}\Bigl(\frac{u}{Q},\mu_{S_\text{thr}}\Bigr)\,
\nonumber\\
&\quad \times 
 \mathscr{S}\Bigl(\boldsymbol{b},\frac{u}{Q}\Bigr)\, D_{q\rightarrow h}(z_{h},\boldsymbol{b})\,
 R_{J}\Bigl[\frac{u}{Q^2},\mu_{J}\rightarrow Q\Bigr]\,
R_{S_{\mathrm{thr}}}\Bigl[\frac{u}{Q},\mu_{S_{\mathrm{thr}}}\rightarrow Q\Bigr]\,
\nonumber\\
&\quad \times 
 R_{{\mathscr{S}}}\Bigl[\boldsymbol{b}, \frac{u}{Q},(\mu_0, \tilde{\zeta}_{\mathscr{S}})\rightarrow \Bigl(Q,\frac{Q^2}{u^2 e^{2\gamma_E}}\Bigr)\Bigr] \,
R[\boldsymbol{b},(\mu_{0}, {\zeta}_{0})\rightarrow (Q,Q^2)]  
\,,
\end{align}
for the hard scale $\mu=Q$.

\subsection{${\cal M}$-scheme}
\label{sec:Mscheme}

In this subsection we explore an alternative resummation scheme that resums the logarithms of $q_T$ and treats the thrust logarithms at fixed-order. 
A similar scheme was proposed in 
\cite{Echevarria:2026vca}.
Nevertheless,  we will see that fitting to \textsc{Pythia} data, the $\mathcal{M}$-scheme even leads to slightly better fits than the other schemes.
It combines the following ingredients at a single scale 
\begin{equation}
    {\cal M}(\boldsymbol{b},u,Q,\mu,\zeta)=J\Bigl(\frac{u}{Q^2},\mu\Bigr)\, 
    S_{\rm thr}\Bigl(\frac{u}{Q},\mu\Bigr)\,
    {\mathscr{S}}\Bigl(\boldsymbol{b},\frac{u}{Q},\mu,\zeta\Bigr)
    \,.
\end{equation}
In this case the  RGEs are
\begin{align}
   \mu\frac{{\rm d}}{{\rm d}\mu}{\cal M}=\gamma^{\cal M}(\mu,\zeta)\,{\cal M}\,, \qquad
   \zeta\frac{{\rm d}}{{\rm d}\zeta}{\cal M}=-{\cal D}(\boldsymbol{b},\mu)\,{\cal M}  \,.
\end{align}
Its anomalous dimension is 
\begin{align}
    \gamma^{\cal M} = \gamma^{J}+\gamma^{S_{\mathrm{thr}}}+\gamma^{\mathscr{S}} = \Gamma_{\mathrm{cusp}}(a_{s})\ln\Bigl(\frac{\mu^{2}}{\zeta}\Bigr)-\gamma^{V}(a_{s})
    = \gamma^F
    \,,
\end{align}
which equals the TMD anomalous dimension,
as required by consistency. While perturbatively $R$ and $R_\mathcal{M}$ are equal, there could be a difference in their nonperturbative contribution. As a starting point, we will take $R=R_\mathcal{M}$.

Since the evolution equations do not depend on $u$, and $\cal{M}$ has the same anomalous dimensions as the TMDs, the $\zeta$-prescription can be applied straightforwardly. For the $e^+e^-$ cross section in region 2 ($\tau Q \ll q_{T} \ll \sqrt\tau Q$) this yields, 
\begin{eqnarray}
\frac{\mathrm{d}\sigma_2}{\mathrm{d}z_{h}\,\mathrm{d}\boldsymbol{q}\,\mathrm{d}\tau}&=&\sum_{q} \sigma_{0,q}(Q) \int \frac{\mathrm{d}\boldsymbol{b}}{(2\pi)^{2}} \int^{\gamma+i\infty}_{\gamma-i\infty} \frac{\mathrm{d}u}{2\pi i} e^{i\boldsymbol{b}\cdot \boldsymbol{q}+u\tau}\,H(Q^2,\mu^2) 
\nn \\ &&\times 
\Bigl(R[\boldsymbol{b},(\mu_{0},\zeta_{0})\rightarrow (Q, Q^2)]\Bigr)^2  {\cal M}(\boldsymbol{b},u)\,  D_{q\rightarrow h}(z_{h},\boldsymbol{b})\,,
\end{eqnarray}
and its ingredients have been defined in secs.~\ref{sec:SIA} and~\ref{sec:indyscheme}.

As in the TMD case, the function $\mathcal{M}(\boldsymbol{b},u)$ is scaleless and fixed at the saddle point, which is ``optimal'' in the sense of ref.~\cite{Scimemi:2018xaf}. 
Decomposing it as a product of a perturbative ($\mathcal{M}^\text{pert}$) and nonperturbative ($\mathcal{M}^\text{NP}$) piece
\begin{align}
    \mathcal{M}=\mathcal{M}^\text{pert}\mathcal{M}^\text{NP}\,,
\end{align}
the perturbative part is up to one-loop order given by
\begin{align}        &\mathcal{M}^\text{pert}=\mathcal{M}^{[0]}+a_s\mathcal{M}^\text{[1]}+\dots \,, \qquad
        \mathcal{M}^{[0]}=1\,, \nonumber\\
        &\mathcal{M}^{[1]}=C_{F}\biggl[ 2\ln^{2}\Bigl(\frac{u}{Q^{2}B^{2}e^{\gamma_{E}}}\Bigr)+3\ln(\frac{u}{Q^{2}B^{2}e^{\gamma_{E}}}) -8\ln^{2} \Bigl( \frac{u}{QB} \Bigr) + 7-\frac{11\pi^{2}}{6}  \biggr]\,.
\end{align}

 We are left to choose the nonperturbative model for $\cal{M}^\text{NP}$ that we discuss in the next section~\ref{sec:Models}. 
A very similar construction holds for the SIDIS case, as the factorization  is obtained from  $e^+e^-$ by replacing the TMD fragmentation function with the TMDPDF, 
\begin{eqnarray}
\frac{\mathrm{d}\sigma^\text{DIS}}{\mathrm{d}x\mathrm{d}\boldsymbol{q}\mathrm{d}\tau}&=&\sum_{q} \sigma_{0,q}(Q) \int \frac{\mathrm{d}\boldsymbol{b}}{(2\pi)^{2}} \int^{\gamma+i\infty}_{\gamma-i\infty} \frac{\mathrm{d}u}{2\pi i}\, e^{i\boldsymbol{b}\cdot \boldsymbol{q}+u\tau}\,  H^\text{DIS}(Q^2,\mu^2)
\nn \\ &&\times 
(R[\boldsymbol{b},(\mu_{0},\zeta_{0})\rightarrow (Q,Q^2)])^2 {\cal M}(\boldsymbol{b},u)  F_{q\leftarrow h}(x,\boldsymbol{b})\,.
\end{eqnarray}
In order to provide predictions for the SIDIS case it is therefore necessary to extract the nonperturbative parts of $\mathcal{M}$ from $e^+e^-$ data.

\section{Nonperturbative modeling}
\label{sec:Models}

Factorization provides a method to identify universal operator matrix elements, and these typically have both a perturbative and nonperturbative part. The nonperturbative part is either obtained using nonperturbative methods or extracted from experiment.
Notably the nonperturbative contributions to the TMD evolution kernel, TMDPDF and TMD fragmentation functions have been extracted using perturbative ingredients at approximate N$^4$LL order in \cite{Moos:2025sal,Bacchetta:2022awv} from Drell-Yan and SIDIS data. These will be used as input in our calculation.
Similar fits have been carried out in the past in refs.~\cite{DAlesio:2014mrz,Scimemi:2017etj,Bacchetta:2017gcc,Bertone:2019nxa,Scimemi:2019cmh,Vladimirov:2019bfa,Bury:2022czx,Moos:2023yfa}. 

The TMD evolution kernel and fragmentation functions have also been obtained from a fit to the TMD distribution of hadrons with respect to the thrust axis~\cite{Belle:2019ywy} at NLL in refs.~\cite{Boglione:2022nzq,Boglione:2023duo}.
While our goal is also to extract the TMDs and evolution kernels, the description of these measurements also requires taking into account the nonperturbative effects for the thrust measurement.

We will denote the nonperturbative part by $\mathcal{M}^\text{NP}$, which appears in addition to the nonperturbative contributions to TMDs, and is the product of three different nonperturbative pieces 
\begin{align}
\mathcal{M}^\text{NP}(b,u)=
J^\text{NP}\Bigl(\frac{u}{Q^2},\mu_0\Bigr)\,
S_\text{thr}^\text{NP}\Bigl(\frac{u}{Q},\mu_0\Bigr)\,{\mathscr{S}}^\text{NP}\Bigl(b, \frac{u}{Q}\Bigr)\,.
\end{align}
In the $\mathcal{M}$-scheme, $\mu_0$ is fixed by the $\zeta$-prescription. In the $u$ and $\tau$-scheme, the jet and soft function would have different natural scales. 

In this paper we take a first step towards a fit with real data, by proposing a parametrization for $\mathcal{M}^\text{NP}$ and testing it using simulated data. 
The specific model we use is
\begin{align} \label{eq:MNP}
    \mathcal{M}^\text{NP}(b,u)= \frac{1}{(1+\mu_1 u)^{\mu_{2}}}
\frac{1}{(1+\mu_{3}\,b\,u)}    \frac{1+\mu_{5}\,b^{\mu_{6}}}{\cosh(\mu_{4}b)}
\,.
\end{align}
The first factor, parameterized by $(\mu_1,\mu_2)$, depends on the thrust measurement through $u$ but not the transverse momentum, and allows us to perform the inverse Laplace transformation numerically, improving the convergence.  The last factor instead just depends on $b$, is very similar to the nonperturbative parametrization of TMDs, and is independent of $u$. The middle factor describes nonperturbative effects intertwining $b$ and $u$. The denominators all lead to damping in the nonperturbative region of phase space, with the relevant scales at which this damping occurs set by $\mu_1, \mu_3$ and $\mu_4$, and the strength of the damping in the first factor controlled by $\mu_2$. The term $\mu_5 b^{\mu_6}$ mimics a correction that has already been found to remove some bias in modeling nonperturbative effects for TMDFFs~\cite{Moos:2025sal}.

More phenomenologically, we expect that $\mu_1$ describes the $\mathcal{O}(\Lambda_\text{QCD}/(\tau Q))$ corrections when $\tau$ becomes small, and similarly $\mu_4$ provides corrections of $\mathcal{O}(\Lambda_\text{QCD}^2/q_T^2)$ when $q_T$ becomes small. Because of the way $u$ and $\boldsymbol{b}$ are defined, we therefore expect $\mu_1 \sim \Lambda_\text{QCD}/Q$, while $\mu_4 \sim \Lambda_\text{QCD}$.
The corrections provided by $\mu_3$ should be relevant when both $q_T$ and $\tau$ are small, $\mathcal{O}(\Lambda_\text{QCD}^2/(\tau Q\, q_T))$, suggesting $\mu_3 \sim \Lambda_\text{QCD}^2/Q$.
Expecting quadratic corrections for small $b$ would imply $\mu_6=2$ such that $\mu_5$ has dimensions of mass squared. However to also fit the sign of this term, we prefer \emph{not} to write it as $(\mu_5 b)^{\mu_6}$.

\subsection{TMD evolution kernel}
\label{sec:Dkernel}

The CS evolution kernel is  determined from the rapidity-divergent part of the TMD soft factor \cite{Echevarria:2015byo,Vladimirov:2020umg}  for small values of $b$. Perturbative calculations have now reached the 4-loop precision~\cite{Li:2016ctv, Vladimirov:2016dll, Duhr:2022yyp, Moult:2022xzt} (see also \cite{Echevarria:2012pw}). The kernel has also a nonperturbative part $\sim  b^2$, as demonstrated by analyses of the renormalon structure \cite{Korchemsky:1994is, Scimemi:2016ffw} and by direct computation \cite{Vladimirov:2020umg}.

Thus the CS kernel can be written as
\begin{eqnarray}\label{def:CS-kernel}
\mathcal{D}(b,\mu)=\mathcal{D}_{\text{pert}}(b^*,\mu^*)+\int_{\mu^*}^\mu \frac{{\rm d}\mu'}{\mu'}\,\Gamma_{\text{cusp}}(\mu')+\mathcal{D}_{\text{NP}}(b),
\end{eqnarray}
where
\begin{eqnarray}\label{CS:scale}
b^*(b)=\frac{b}{\sqrt{1+\frac{b^2}{B^2_{\text{NP}}}}},\qquad \mu^*(b)=\frac{2e^{-\gamma_E}}{b^*(b)},\qquad B_{\text{NP}}=1.5\text{ GeV}^{-1}\,.
\end{eqnarray}
$\mathcal{D}_{\text{pert}}$ is the perturbative part of the kernel, and the nonperturbative part $\mathcal{D}_{\text{NP}}$ is modeled by 
\begin{eqnarray}\label{CS:NP-part}
\mathcal{D}_{\text{NP}}(b)=b\,b^*\biggl[c_0+c_1\ln \Bigl(\frac{b^*}{B_{\text{NP}}}\Bigr)\biggr],
\end{eqnarray}
with $c_0$ and $c_1$ free parameters extracted from the fit of Drell-Yan and SIDIS data~\cite{Moos:2025sal},
\begin{align}
    c_0&=  0.0859^{+0.0023}_{-0.0017}\text{ GeV$^{-2}$ }\,, \qquad
    c_1 =  0.0303^{0.0038}_{-0.0041}\text{ GeV$^{-2}$ }\,.
\end{align}

\subsection{TMD fragmentation function }

In the rest of the paper and in the numerical analysis we restrict ourselves to the unpolarized TMDFF (indicated conventionally by the subscript 1),
which can be decomposed as
\begin{align}
\label{def:TMDFF}
D_{1,f\to h}(z,b)&=\sum_{f'}\mathbb{C}_{f\to f'}(z,b^{*(\text{FF})}_{\text{OPE}},\mu^{\text{FF}}_{\text{OPE}})\otimes d_{1,f'\to h}(z,\mu^{\text{FF}}_{\text{OPE}})\,d_{\text{NP}}^{f/h}(z,b).
\end{align}
Here $d_1$ is the unpolarized collinear fragmentation function, $\mathbb{C}$ are the matching coefficients and $d_{\text{NP}}$ parametrizes the nonperturbative large-$b$ behavior, and the symbol $\otimes$ indicates a Mellin convolution. 
In \texttt{artemide}
$\mathbb{C}$ is evaluated at N$^3$LO accuracy~\cite{Luo:2019hmp, Ebert:2020qef} and the large-$z$ asymptotic is resummed followed the approach of ref.~\cite{delRio:2025qgz}.  

The scale of this operator product expansion for the TMDFF is $\mu^{\text{FF}}_{\text{OPE}}$, which is independent from other scales in the factorization. The parameter $b^*_{\text{OPE}}$ is also independent from similar ones appearing in the factorization. The expressions we use for them are
\begin{align}\label{muOPE:FF}
\mu_{\text{OPE}}^{\text{FF}}&
=\frac{2e^{-\gamma_E}\,z}{b}+5\text{ GeV},
\\\label{bOPE:FF}
b^{*(\text{FF})}_{\text{OPE}}&=b\,e^{-a b^2}+\frac{2e^{-\gamma_E}}{\mu^{\text{FF}}_{\text{OPE}}(b)}\,\bigl(1-e^{-ab^2}\bigr),
\end{align}
with the value of $a=0.04$ GeV$^2$ fixed. 

The nonperturbative flavor-dependent part $d_{\text{NP}}^f(z,b)$ is defined as
\begin{eqnarray}\label{d1:np-part}
d^{f/h}_{\text{NP}}(z,b) = \frac{1+\eta_1^{h,f}\frac{b^2}{z^2}}{\cosh\(\eta_0^h \frac{b}{z}\)}\,.
\end{eqnarray}
The values of these parameters are illustrated for $\pi^+$ in tab.~\ref{tab:etapi}, for which the
valence $u$, $\bar d$ flavors, $\bar u$ flavor, and  remaining flavors $r$ are separated. 

\begin{table}[]
    \centering
    \begin{tabular}{||c|c|c|c|c||}
    \hline
       $\eta^{\pi^+}_{0}$ & $\eta^{\pi^+, u}_{1}$ & $\eta^{\pi, \bar{d}}_{1}$ & $\eta^{\pi^+, \bar{u}}_{1}$ & $\eta^{\pi^+, r}_{1}$ \\ \hline
        0.696 & 0.626 & 0.003 & 0.61 & -0.47 \\
         \hline
    \end{tabular}
    \caption{\label{tab:etapi}
    The parameters for TMDFF for $\pi^+$ in ART25. See eq.~\eqref{d1:np-part} for the parametrization.}
    \label{tab:ART25TMDFFparameters}
\end{table}

\subsection{Inverse Laplace transform}
\label{sec:laplace}

The final cross section is obtained by first performing the inverse Laplace transform of  $u$ numerically, and then the Fourier transform of $b$. The convergence of these numerical methods depend on the nonperturbative model, which is why we discuss it here. The inverse Laplace transform uses Talbot's method, which is based on the following deformation of the Bromwich contour: the idea is to  deform the contour from running parallel to the imaginary axis to one that opens toward the negative real axis,
\begin{eqnarray}
    &&u=\sigma+\frac{M}{\tau}(\theta\cot(\theta)+i\theta)\,,
\end{eqnarray}
with $-\pi < \theta < \pi$.
The values of $\sigma= 0.01$ and $M= 1 + 20 \tau$ have been tested (to a precision of 11 decimal places) with a toy model similar to the actual function which we need to integrate.
For very small values of $\tau$ ($\lesssim 0.03$) the numerical inverse Laplace transform becomes less stable, since the damping from $\exp(u\tau)$ becomes small compared to the oscillation coming from the logarithms of $u$. Thus we restrict our study to $\tau > 0.03$.

\section{Results}
\label{sec:results}

\begin{figure} [b]
    \centering
    \includegraphics[width=0.65\linewidth]{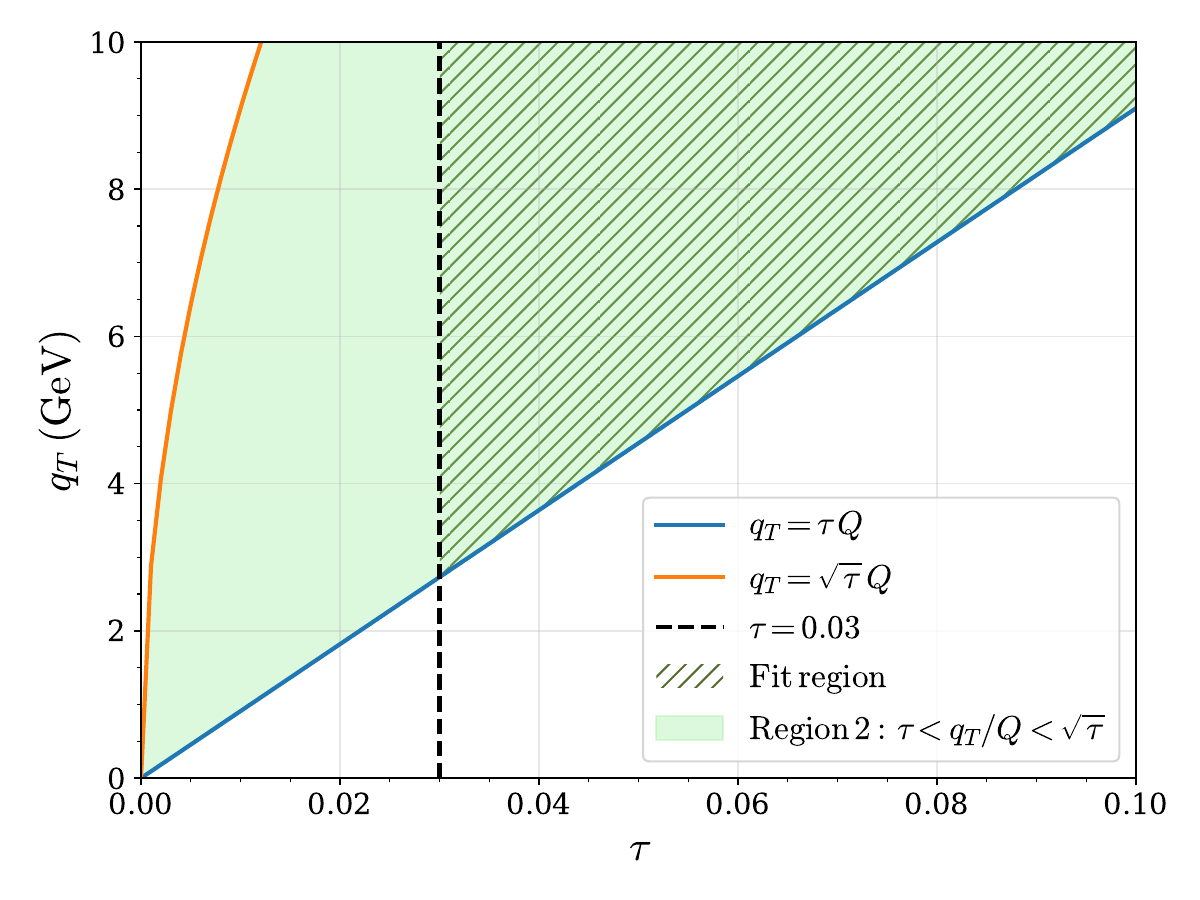}
    \caption{Shown are the full region 2 of the phase space and the region considered for the fit.}
    \label{fig:region2}
\end{figure}

In order to test our formalism, both the perturbative predictions as well as the nonperturbative modeling, we compare to 10 million events generated by \textsc{Pythia}~8.3 at $Q=91.0$ GeV. 
We estimate the systematic uncertainty to be 15\% in each bin, treating bins as uncorrelated. We add this in quadrature to the statistical uncertainty from the Monte Carlo data. 
We do not take the theory uncertainties into account in the fit, but will discuss them at the end of this section.
For this test we have selected 10 bins in $\tau$ ranging from 0 to 0.1, 50 bins in $q_T$ from $0$ to $10$ GeV, and 20 bins in $z$ from 0 to 1. 
After restricting to region 2 and imposing $\tau>0.03$, see fig.~\ref{fig:region2}, we have a total of 1788 data points for our analysis.

We have resummed the logarithms of $q_T/Q$ and $\tau$, implementing the $\zeta$-prescription, and perform a numerical inverse Laplace transform from $u$- to $\tau$-space. We recap the following three choices
for the resummation of logarithms of $\tau$, discussed in sec.~\ref{sec:resummation}: 
\begin{itemize}
    \item $\tau$-scheme: choose $\mu$-scales in $\tau$-space, 
    \item $u$-scheme: choose $\mu$-scales in $u$-space, except for the scale of $\alpha_s$, which is instead taken in $\tau$-space, 
    \item $\mathcal{M}$-scheme: turn off the resummation of $\tau$.
\end{itemize}

\begin{table}[t]
    \centering
    \resizebox{\columnwidth}{!}{
    \begin{tabular}{|c||c|c|c|c|c|c|c|}
        \hline
         & $\mu_{1}$ & $\mu_{2}$ & $\mu_{3}$ & $\mu_{4}$ & $\mu_{5}$ & $\mu_6$ & $\chi^{2}/N$  \\
         \hline \hline
   $\tau$-scheme &  0.00458 & 1.61  & 0.0397 & 0.00239 & -9.96  &  0.161 & 2.57   \\
        \hline
        $u$-scheme &  0.0285 & 0.319 & 0.0160  & 2.68 &  -2.77 &  2.32 & 1.71 \\
        \hline
         $\mathcal{M}$-scheme &  0.0120 & 2.19 &  6.30 $\cdot 10^{-8}$ & 0.00682 & -7.03  & 2.94 & 1.69  
         \\ \hline\hline
    \end{tabular}
   }  
    \caption{Results for the fit parameters of the nonperturbative model in eq.~\eqref{eq:MNP} for $e^+e^-$ in region 2, with
    $\tau > 0.03$ and $ 0.2 < z < 0.8$. The rows correspond to the results for the three resummation schemes discussed in secs.~\ref{sec:indyscheme} and~\ref{sec:Mscheme}. Dimensionful parameters are in the appropriate units of GeV. We use 1788 points generated by \textsc{Pythia}~8.3 at $Q=91.0$ GeV. 
    }
    \label{tab:Fits}
\end{table}

\begin{figure}
    \centering
    \includegraphics[width=0.45\linewidth]{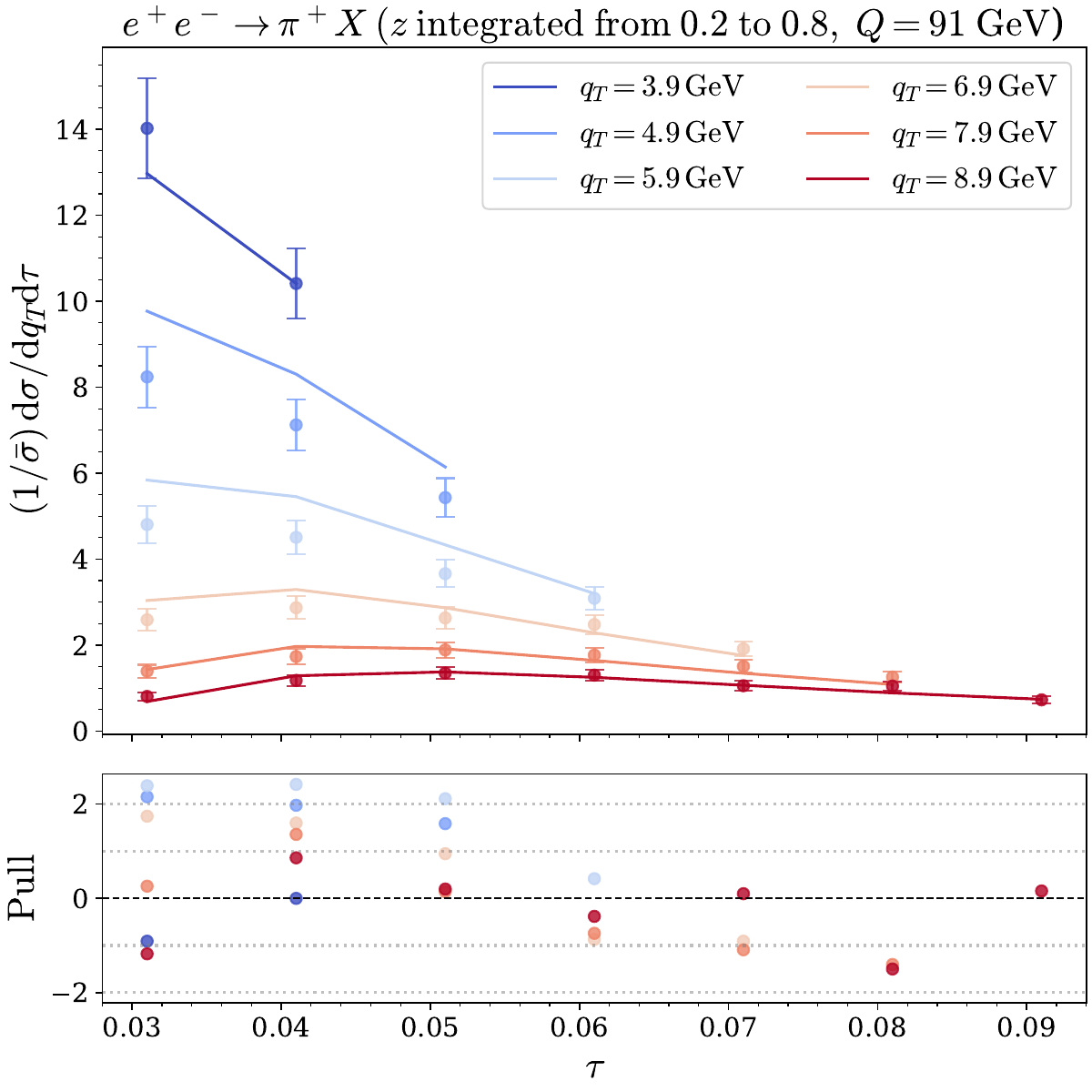}
    \includegraphics[width=0.45\linewidth]{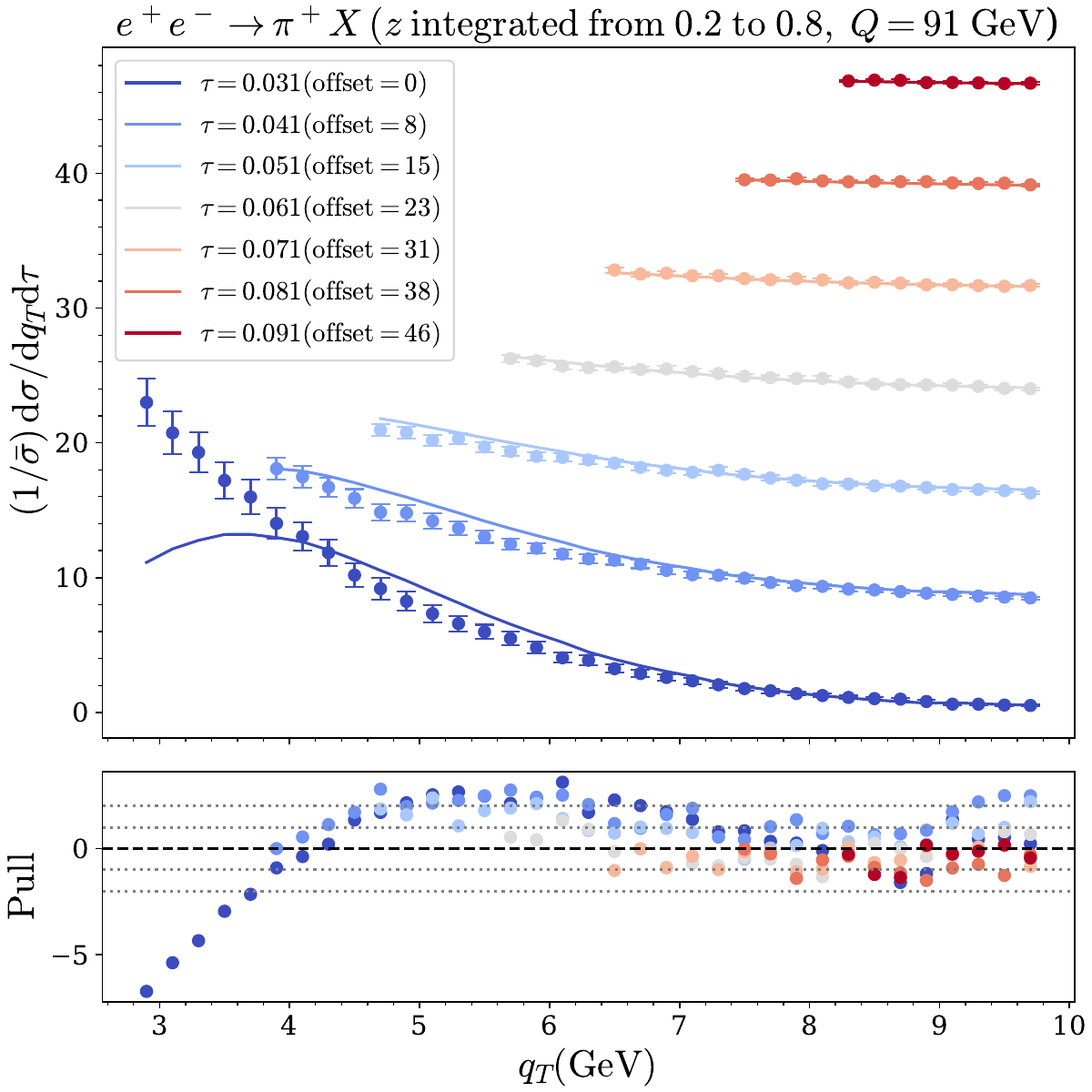}
    \caption{The normalized cross section of eq.~(\ref{eq:normsigma}) as a function of $q_T$ (left) and $\tau$ (right). Shown are the \textsc{Pythia}~8.3 data with statistical uncertainties, as well as the fit result. We include an offset (of indicated size) to space out the different curves.
    The pull is shown in the lower panel.}
    \label{fig:qt-tauplot}
\end{figure}

\begin{figure}
    \centering
    \includegraphics[width=0.45\linewidth]{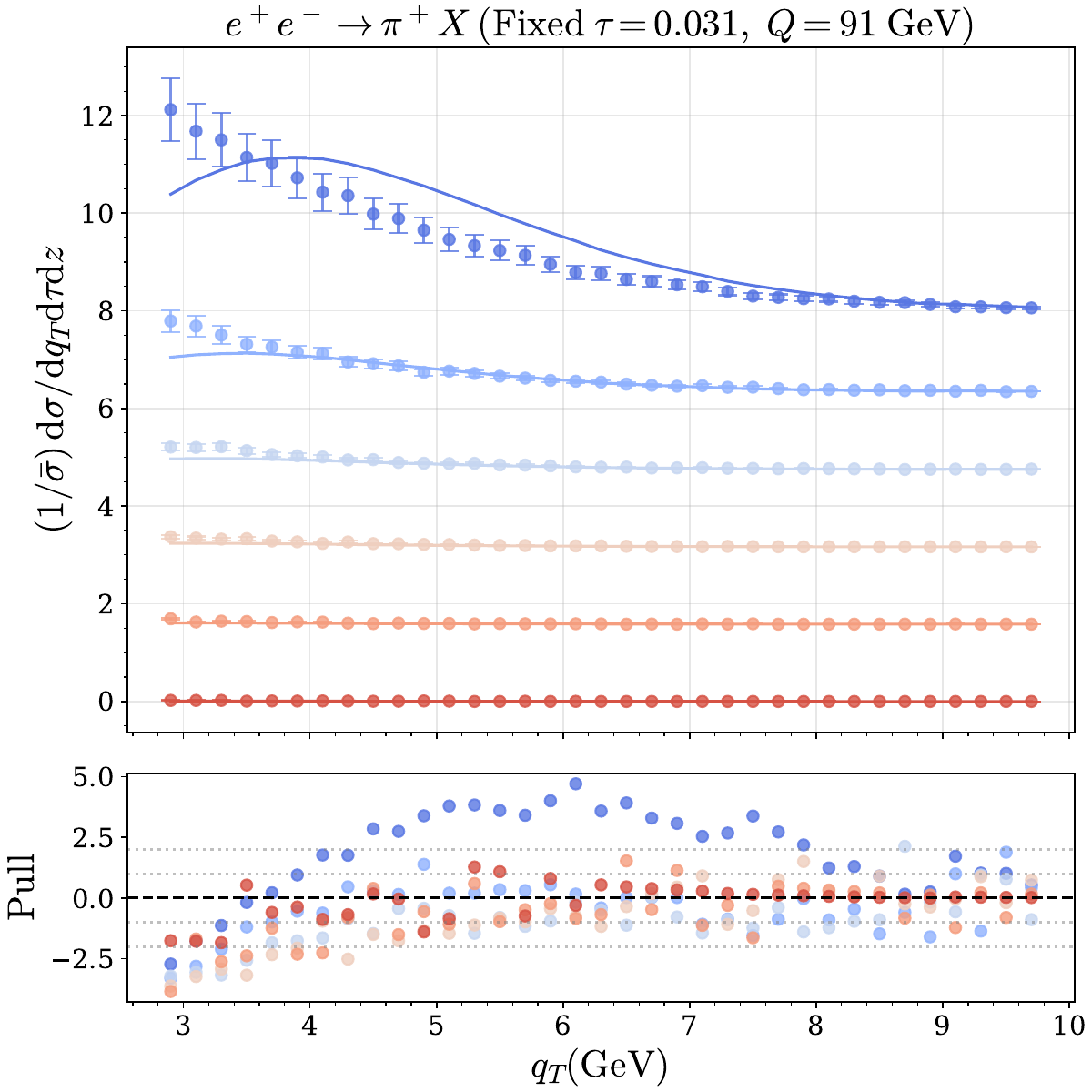}
     \includegraphics[width=0.45\linewidth]{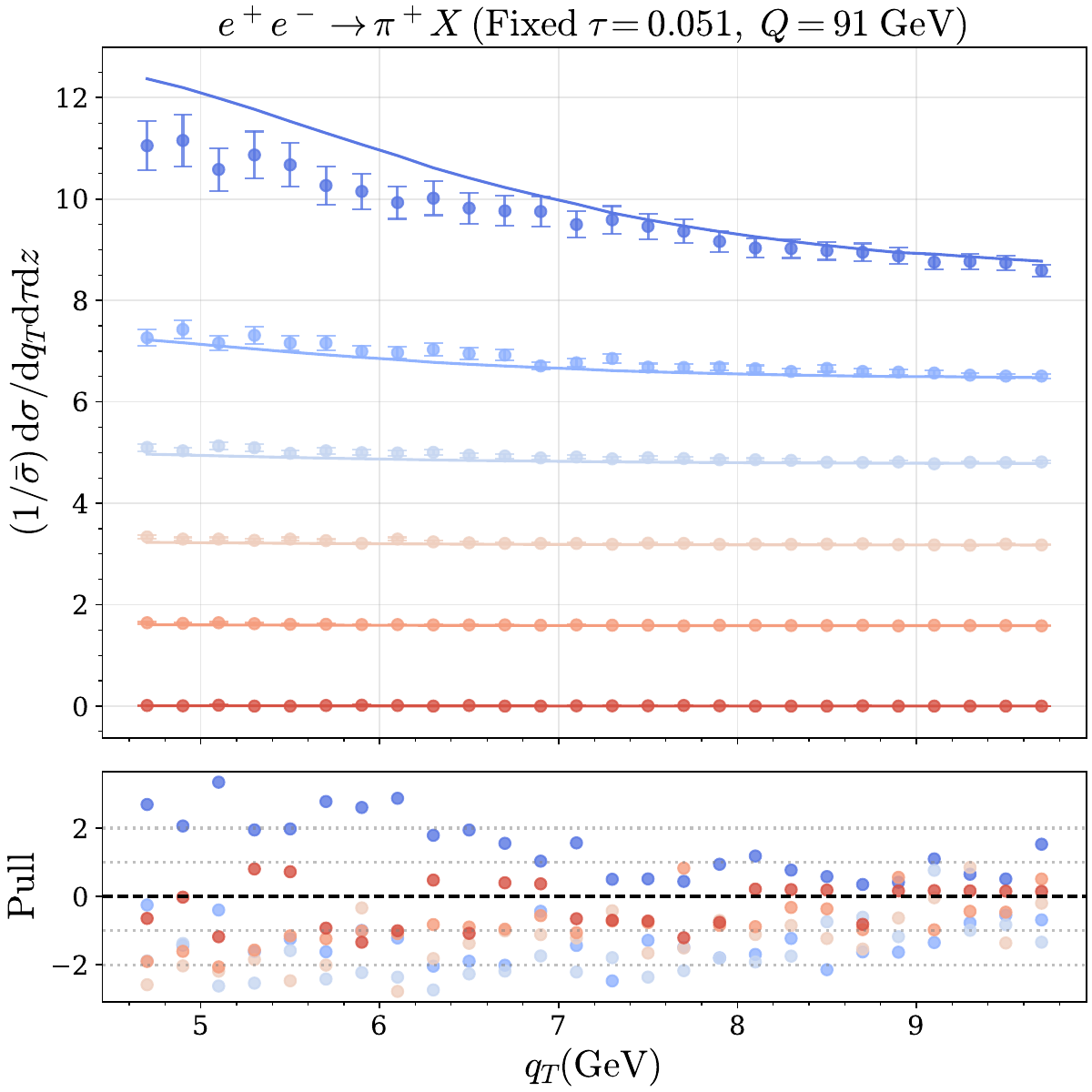}
       \includegraphics[width=0.45\linewidth]{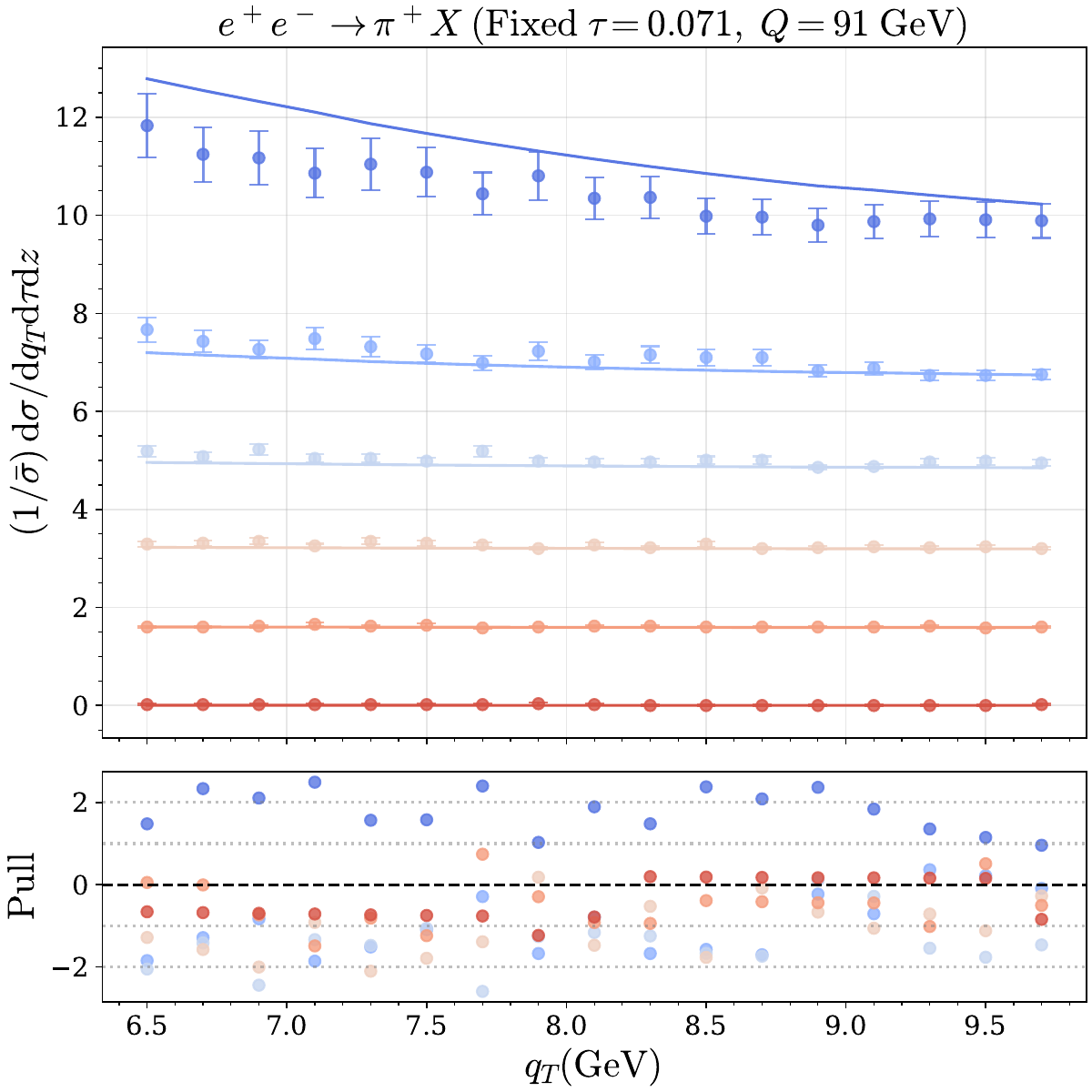}
      \includegraphics[width=0.45\linewidth]{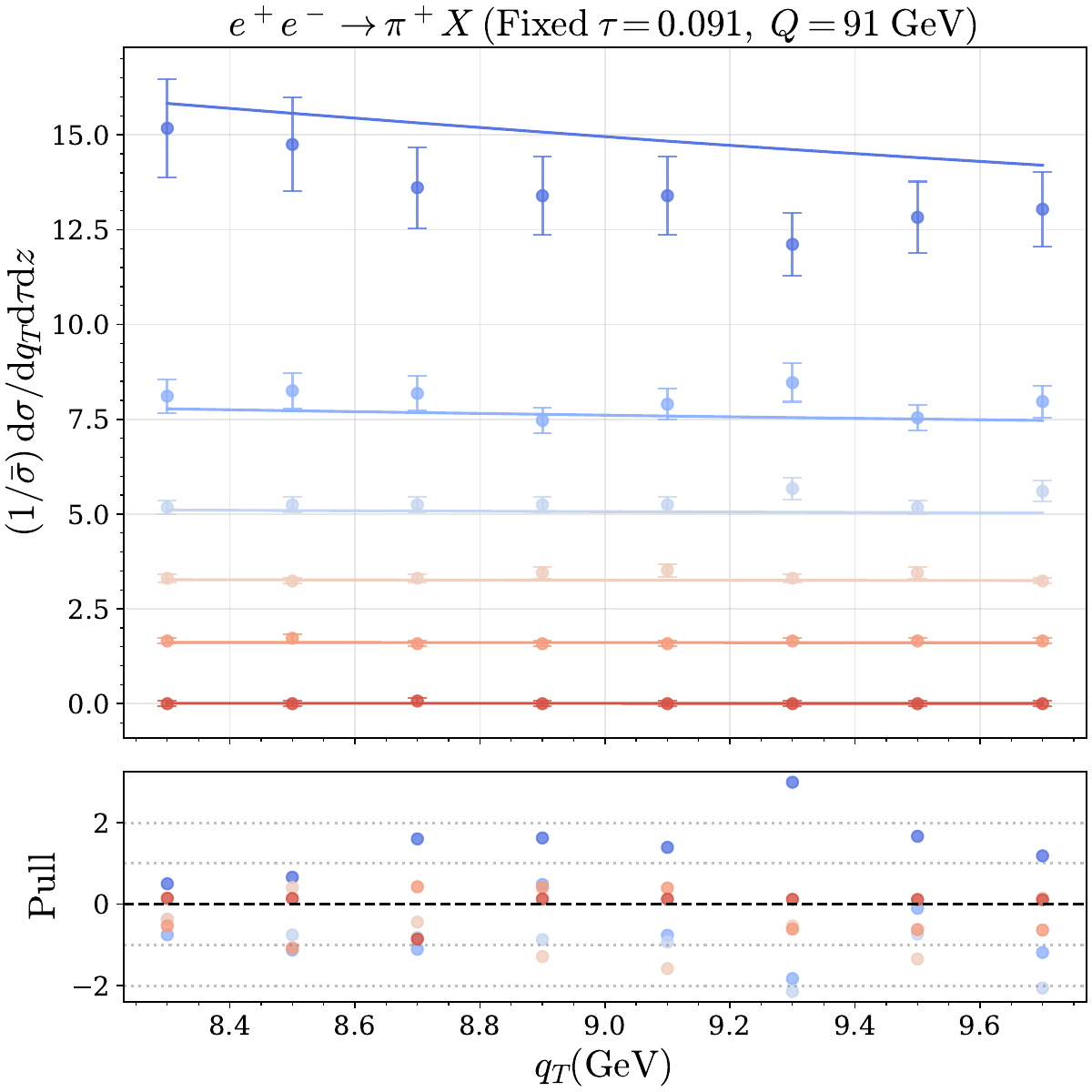}
      \includegraphics[width=0.25\linewidth]{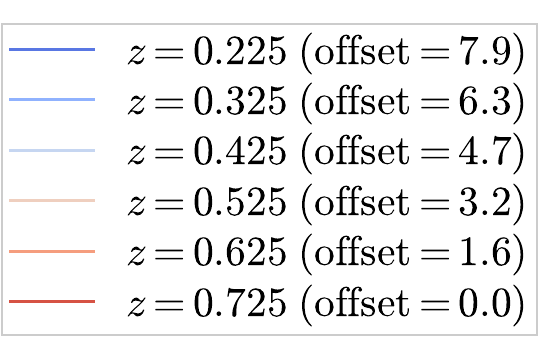}
    \caption{The normalized cross section of eq.~(\ref{eq:normsigma2}) as a function of $q_T$ and several choices of $z$ and $\tau$. Shown are the \textsc{Pythia}~8.3 data with statistical uncertainties, as well as the fit result. The pull is shown in the lower panel.}
    \label{fig:zqt-tauplot}
\end{figure}

\begin{figure}
    \centering
    \includegraphics[width=0.45\linewidth]{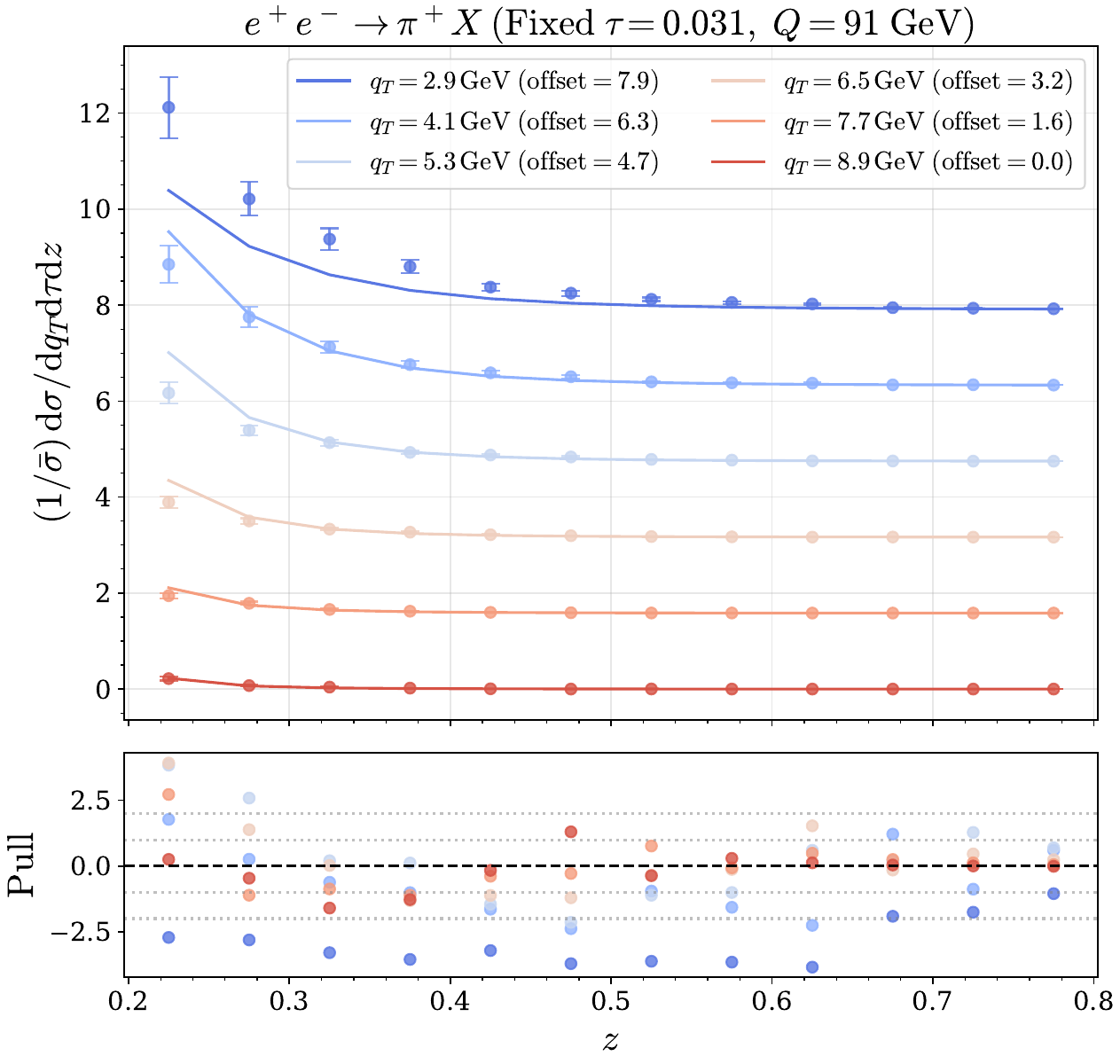}
     \includegraphics[width=0.45\linewidth]{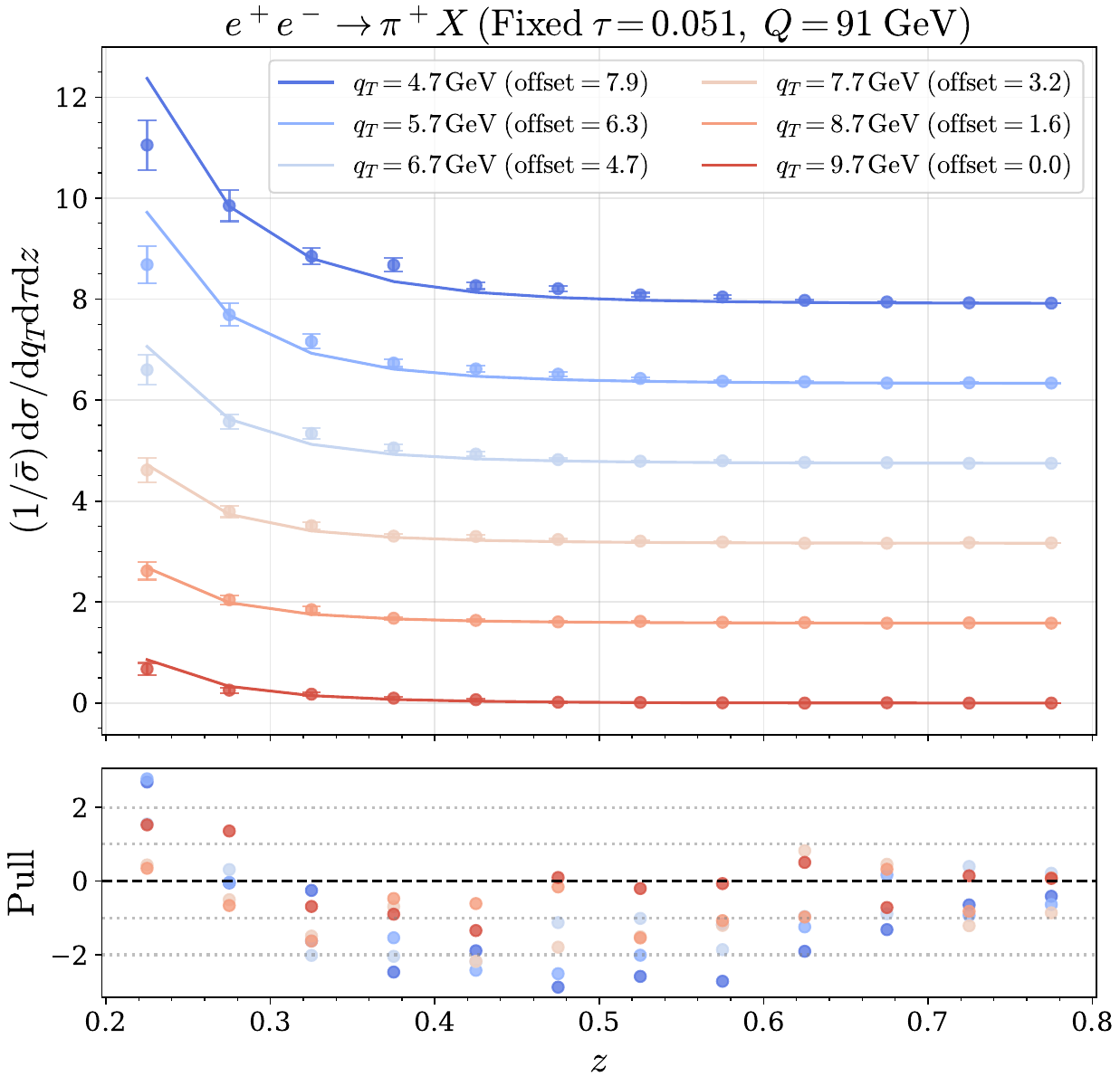}
    \includegraphics[width=0.45\linewidth]{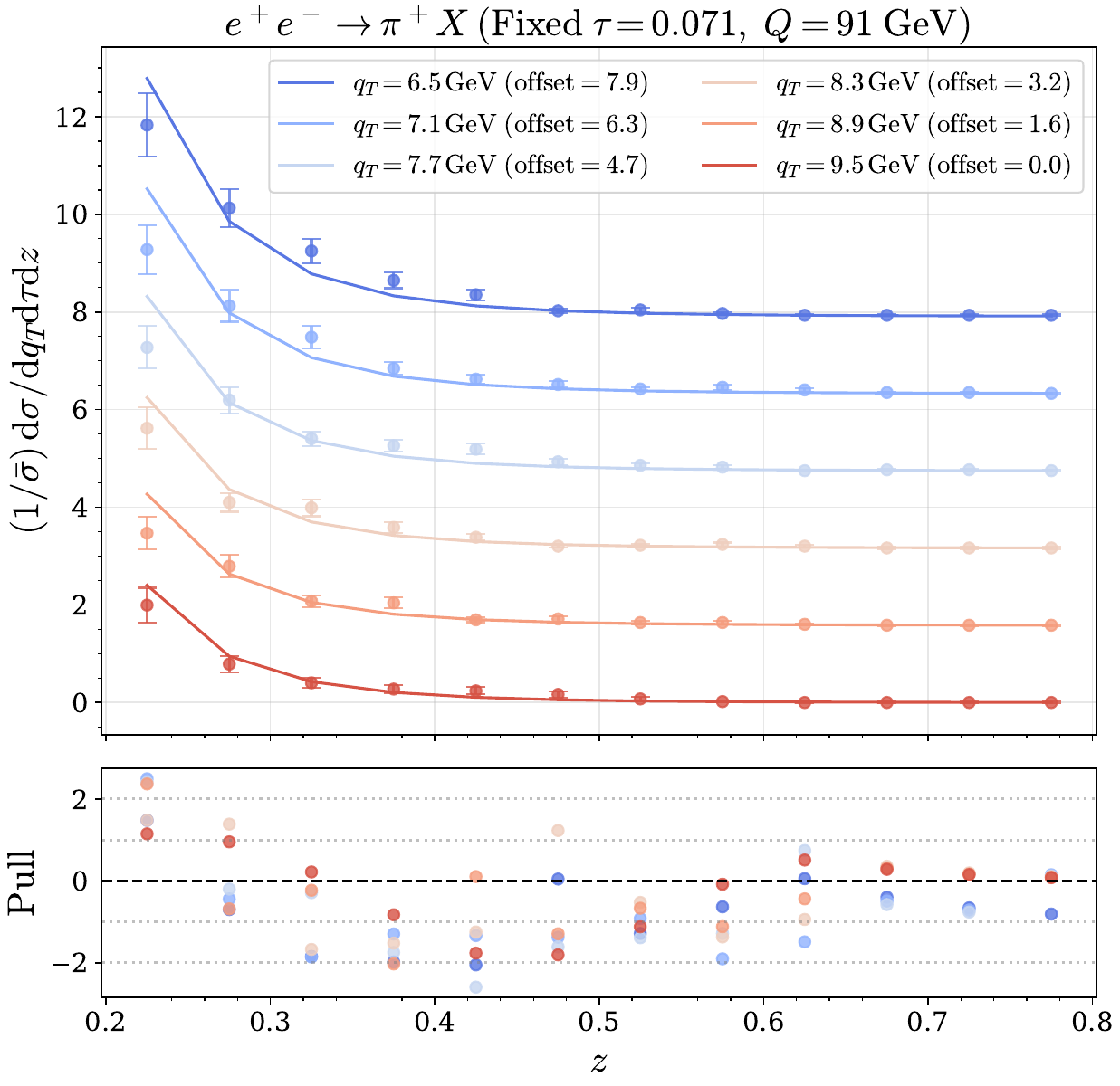}
     \includegraphics[width=0.45\linewidth]{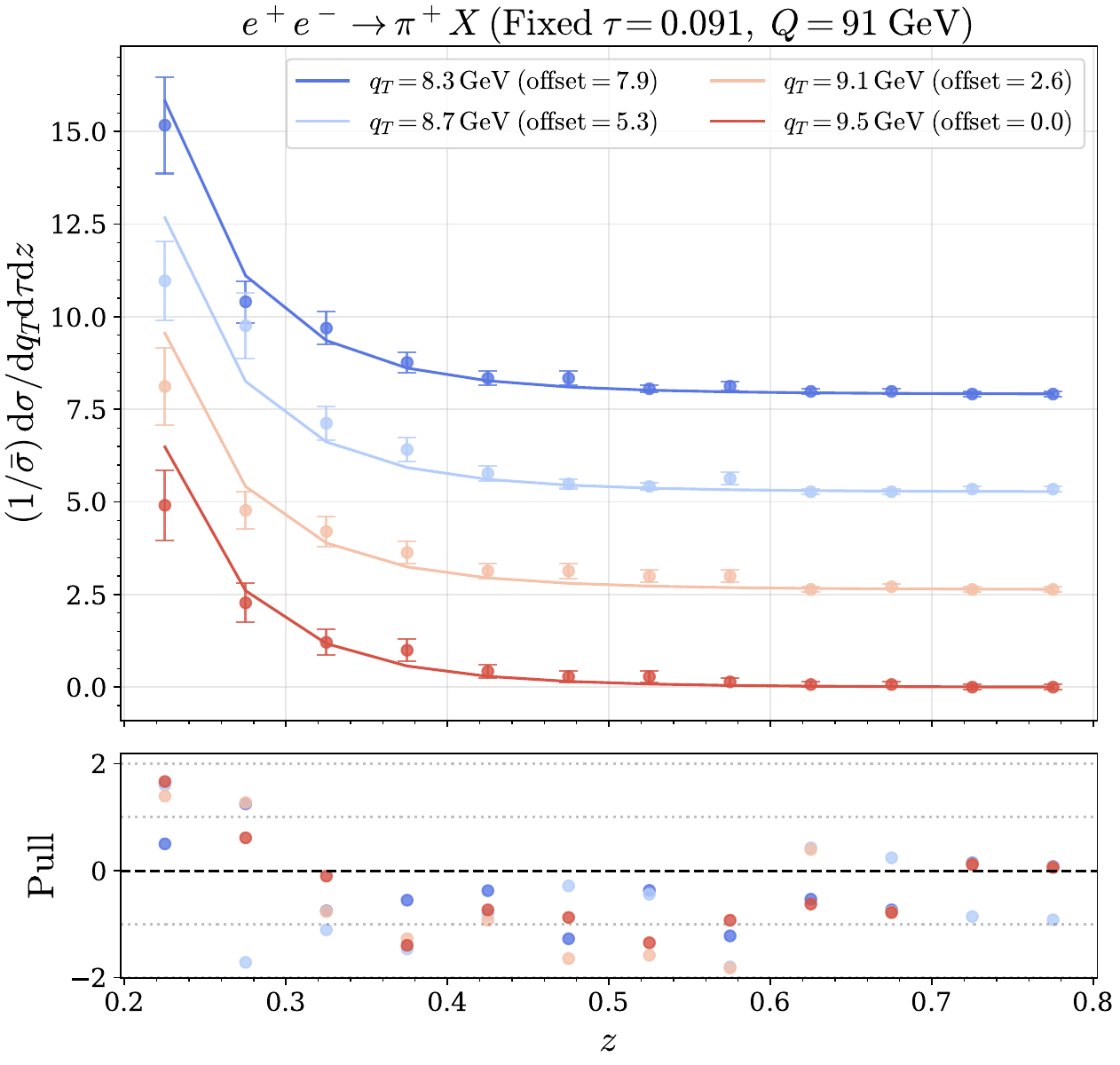}
    \caption{Same as fig.~\ref{fig:zqt-tauplot}, but as function of $z$ for several choices of $q_T$.}
    \label{fig:zqt-plot}
\end{figure}

We have performed a fit of the nonperturbative model $\mathcal{M}^{\rm NP}$,  using the TMDFF extracted from ART25 as input. This is not a real determination of these parameters, for which we need real data, but tests the validity of our framework and the flexibility of our nonperturbative model.
The results are reported in tab.~\ref{tab:Fits}. From the obtained $\chi^2/N$ we observe that the fit of the $u$- and $\mathcal{M}$-schemes is better than for the $\tau$-scheme. 
Concerning the values of the nonperturbative parameters, we observe that $\mu_1 \sim \Lambda_\text{QCD}/Q$, as expected from the analysis in sec.~\ref{sec:Models}. Similarly, for the $\tau$-scheme and $u$-scheme we have $\mu_3 \sim \Lambda_\text{QCD}^2/Q$, in line with expectations. Since $\mu_3$ is the unique nonperturbative parameter correlating $q_T$ and $\tau$, the unusual small value for $\mu_3$ in the $\mathcal{M}$-scheme suggests all correlations are already captured by perturbation theory in this scheme. The factor involving $\mu_4$ is also present in the TMDFF, see eq.~\eqref{d1:np-part}, and for the $\tau$-scheme and $\mathcal{M}$-scheme the smallness of $\mu_4$ suggests no additional suppression is needed for large $b$. We expect $\mu_6 = 2$ because the leading corrections for TMDs are quadratic, which is roughly the case for the $u$-scheme and $\mathcal{M}$-scheme. We expect the other parameters to be order 1, in the appropriate GeV units, which is borne out by the fit. All parameters do differ significantly between the various schemes, highlighting that the meaning of these nonperturbative parameters is intimately connected to the perturbative predictions.

In order to understand the obtained values of $\chi^2/N$
we  plot  the cross section, projecting the triple differential cross section in various ways. First, in fig.~\ref{fig:qt-tauplot} we show the cross section integrated over $z$,
\begin{align}
    \frac{1}{\bar \sigma}\int_{0.2}^{0.8} \!{\rm d}z\;\;\frac{{\rm d}\sigma}{{\rm d}q_T\,{\rm d}\tau\, {\rm d}z} \quad \text{with} \quad \bar\sigma =\int_\text{Region 2}
    \!{\rm d}\tau\,
     {\rm d}q_T\,
    \int_{0.2}^{0.8} \!{\rm d}z\;\;\frac{{\rm d}\sigma}{{\rm d}q_T\,{\rm d}\tau\, {\rm d}z}\,.
    \label{eq:normsigma}
\end{align}
We show the $q_T$ distribution for slices in $\tau$ and vice versa, using the $\mathcal{M}$-scheme. Both plots show that there are deviations from fits for low values of $\tau\lesssim 0.04$. In this region a more detailed analysis, including both higher-order contributions and nonperturbative effects, would be required.

In fig.~\ref{fig:zqt-tauplot} and \ref{fig:zqt-plot} we show 
\begin{align}
    \frac{1}{\bar \sigma(\tau)} \frac{{\rm d} \sigma}{{\rm d}q_T\,{\rm d}\tau\, {\rm d}z} \quad \text{with} \quad \bar\sigma(\tau) =
    \int_\text{Region 2}
    \! {\rm d}q_T
    \int_{0.2}^{0.8} \!{\rm d}z\;\;\frac{{\rm d}\sigma}{{\rm d}q_T\,{\rm d}\tau\, {\rm d}z}\,,
    \label{eq:normsigma2}
\end{align}
as function of $q_T$ for slices in $z$ and vice versa, for fixed values of $\tau$. The cross section is now normalized for each individual value of $\tau$. The agreement is decent, except for small values of $z$ and $\tau$. We also note that the $z$-dependence is entirely due to the TMD fragmentation function, suggesting this kind of data may be relevant to fit the TMDFF for low values of $z$.

\begin{figure}
    \centering    
    \includegraphics[width=0.45\linewidth]{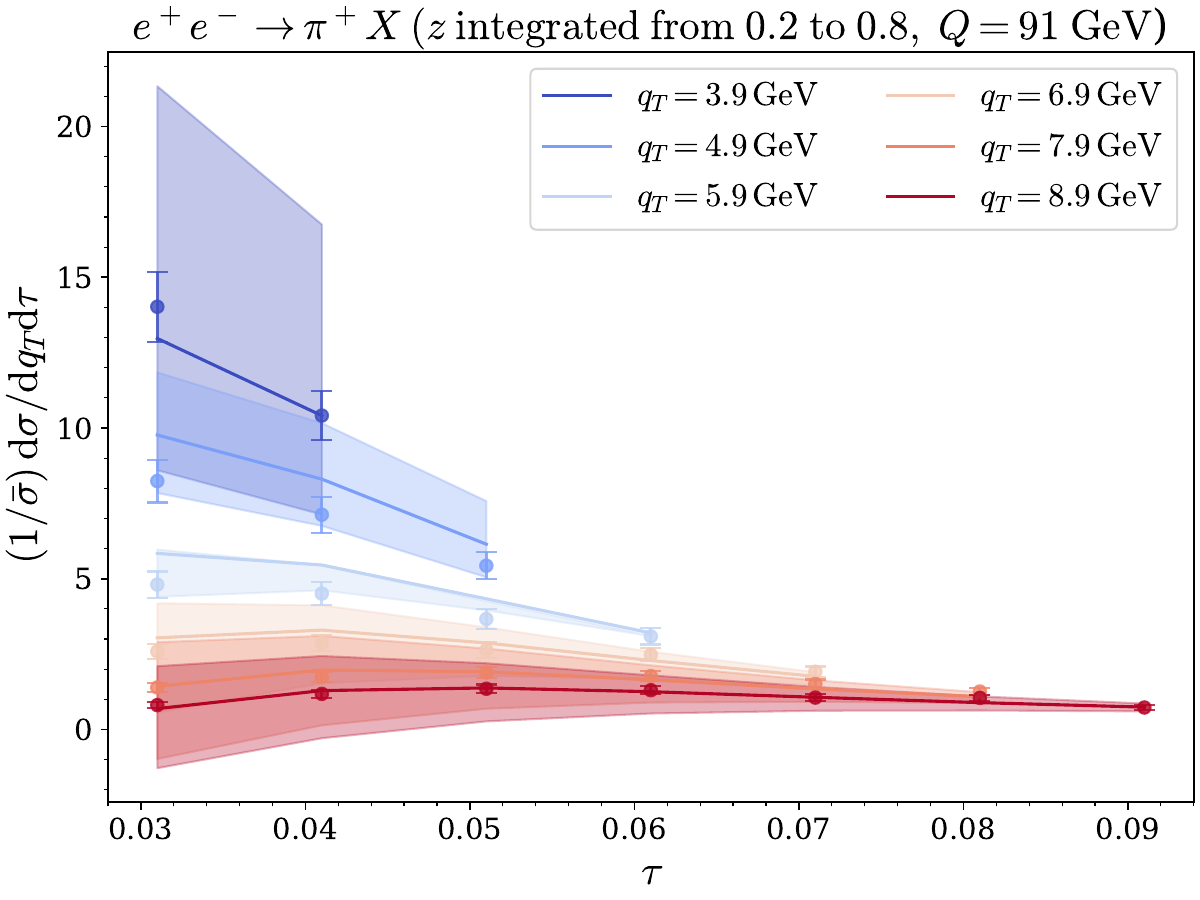}
    \includegraphics[width=0.45\linewidth]{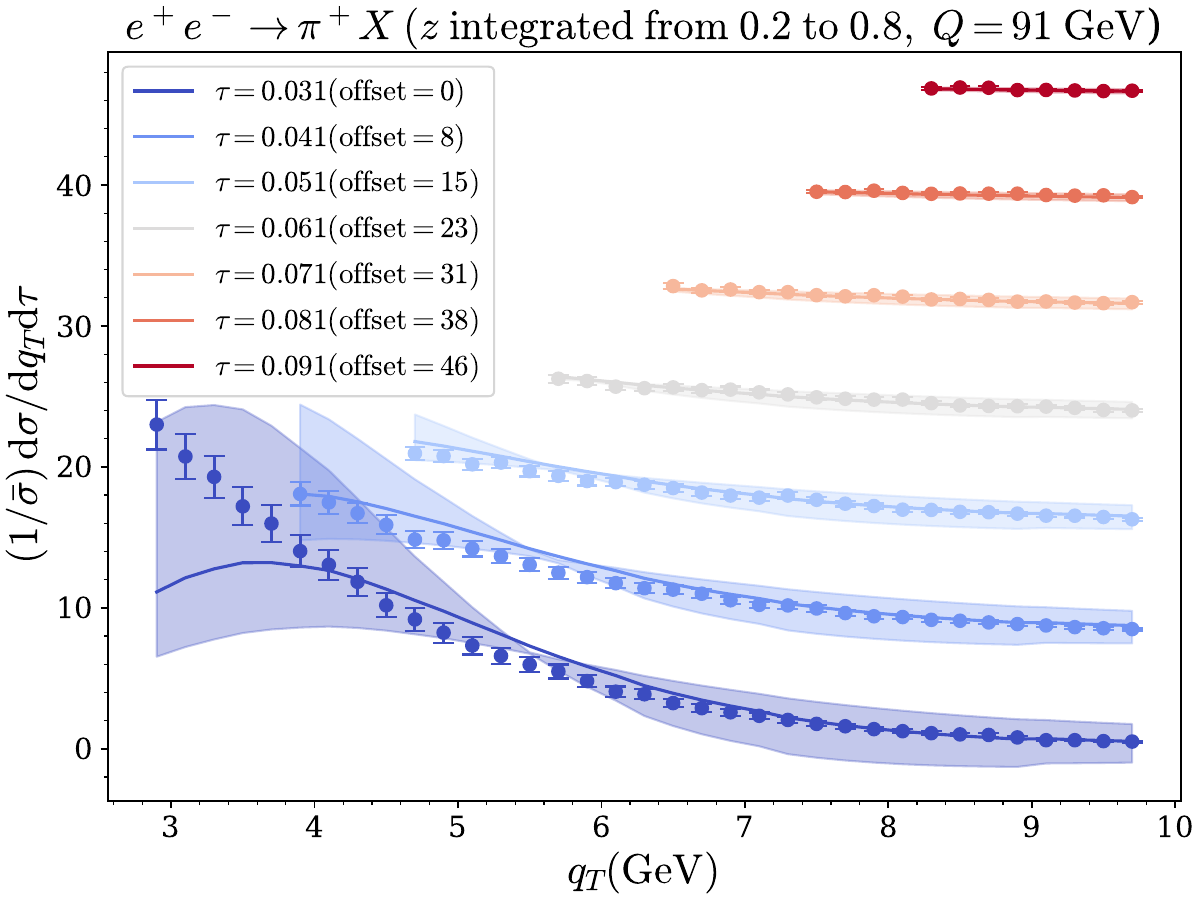}
    \caption{Same as fig.~\ref{fig:qt-tauplot}, including the perturbative uncertainty band from the hard scale variation, $\mu\in (2,1/2)Q$.}
    \label{fig:hardscale}
\end{figure}

Finally  in fig.~\ref{fig:hardscale} we plot the perturbative uncertainty band that come from varying the hard scale in the interval $\mu\in (2,1/2)Q$. This is the generally the largest theory error, though there are also contributions from the fragmentation functions and CS-kernel inherited from ART25.
The band from the scale variation contains the \textsc{Pythia} data, even  when $\tau$ and $q_T$ are small, so if we had included theory errors in the fit we would have obtained a much lower value of $\chi^2/N$.
The size of the bands also identify regions of phase space where more care in the treatment of perturbative and nonperturbative effects may be needed.
While this analysis uses Monte Carlo data, it shows that our resummed predictions and proposed nonperturbative model work.
At the same time the conclusions one can draw from this are limited, and the real test will come from analyzing experimental data.

\section{Conclusions}
\label{sec:conclusions}
In this paper we have examined
the factorization theorems  for the measurement of a TMD with respect to the thrust axis in $e^+e^-$ and 1-jettiness in SIDIS. In $e^+e^-$
the factorization proposed in the literature \cite{Makris:2020ltr,Boglione:2020auc} has been completed by giving the operator definition of the soft function $S_\text{hemi}$, and a one-loop check using the $\delta$-regulator has been performed.
We then proceeded to discuss how a similar measurement can be performed in semi-inclusive deep inelastic scattering (SIDIS). In this case we considered (two possibilities for) the 1-jettiness axis, and obtained the corresponding factorization.

We have tested the $e^+e^-$ case for 
$\tau Q\ll q_T\ll \sqrt{\tau} Q$,
providing a  NNLL resummation,  discussing  nonperturbative effects and testing the framework by fitting to \textsc{Pythia}~8.3  data  at $Q=91.0$ GeV. 
The fit with our nonperturbative model is consistent with \textsc{Pythia} data, validating the structure of the model we adopted. The largest differences arise at small values of $z$, $\tau$ and $q_T$. Here the perturbative uncertainties are also larger, requiring improved perturbative precision before drawing final conclusions. The difference at small $z$ may indicate that the used TMDFF should be corrected there.

We have tested several resummation schemes.
The more relevant ones are the $u$-scheme and $\mathcal{M}$-scheme.
In the former the resummations of logarithms of $\tau Q/q_T$ is carried out, while in latter it is not.
As an advantage the $\mathcal{M}$-scheme  is easier to implement, because it simply inherits the TMD resummation.
 Surprisingly the two schemes give very similar $\chi^2/N$, but some nonperturbative parameters are very different.
In particular, in the $\mathcal{M}$-scheme, the nonperturbative correlations between $q_T$ and $\tau$ are negligible. 
Improving the perturbative precision and comparing with real data may clarify this aspect.

An important step in this direction is the full two-loop calculation of the collinear-soft function.

A natural extension of this project is to apply the present formalism to fit the Belle 
data~\cite{Belle:2019ywy}, and make predictions for the forthcoming Electron Ion Collider. Finally, the present study is also interesting for its possible application to the FCC-ee and a reanalysis of LEP data.

\acknowledgments

We thank  A.~Vladimirov  for technical discussions on \texttt{artemide}. P.A.G.G.~is supported by the Ministry of Education contract FPI-PRE2020-094385.
D.D.F.~is supported by a contract provided  by \textit{Atracci\'on de Talento Investigador} program of the Comunidad de Madrid (Spain) No. 2022-T1/TIC-24024. 
This project is supported by the Spanish Ministerio de Ciencias y Innovaci\'on Grant No. PID2022-136510NB-C31 funded by MCIN/AEI/ 10.13039/501100011033. This project has received funding from the European Union Horizon research Marie Skłodowska-Curie Actions – Staff Exchanges, HORIZON-MSCA-2023-SE-01-101182937-HeI and COST action 24159 (SHARP).
I.S.~thanks NIKHEF for its kind hospitality during  the stages supported by the HeI project.

\appendix
\section{Appendix}
\label{sec:appendixa}

\subsection{Wilson lines}

The Wilson lines entering the definition of the soft functions $S_{\rm thr}$ and $S_{\rm hemi}$ are
\begin{align}
\label{eq:S_SIA}
S_{n}(0)=\boldsymbol{P}\exp\biggl[ -ig\int^{s \infty}_{0} {\rm d}s\; n\cdot A_{s}(sn)\biggr]  \,, \quad 
S_{\bar{n}}(0)=\boldsymbol{P}\exp\biggl[- ig\int_{0}^{s \infty} {\rm d}s\; \bar{n}\cdot A_{s}(s\bar{n})\biggr]\,,
\end{align}
with $s=+1$ for final states and $s=-1$ for initial states.
Similarly, the Wilson lines for the collinear-soft function are~\cite{Bauer:2011uc}
\begin{align} \label{eq:VX}
X_{n}(0)=\boldsymbol{P}\exp\biggl[ -ig\int^{\infty}_{0} {\rm d}s\; n\cdot A_{ncs}(sn)\biggr]  \,, \quad V_{n}(0)=\boldsymbol{P}\exp\biggl[ -ig\int^{\infty}_{0} {\rm d}s\; \bar{n}\cdot A_{ncs}(s\bar{n})\biggr]\,.\;
\end{align}

\subsection{Evolution factors}

The $S$ and $A$ functions that enter in the solutions to the RGEs are given by
\begin{align}
 S(\nu,\mu)&=-\int_{\alpha_{s}(\nu)}^{\alpha_{s}(\mu)} \mathrm{d}\alpha\, \frac{\Gamma_{\mathrm{cusp}}(\alpha)}{\beta(\alpha)} \int_{\alpha_{s}(\nu)}^{\alpha} \frac{\mathrm{d}\alpha'}{\beta(\alpha')}
 \,,\nn
\\
     A_{\Gamma}(\nu,\mu)&=-\int_{\alpha_{s}(\nu)}^{\alpha_{s}(\mu)} \mathrm{d}\alpha\, \frac{\Gamma_{\mathrm{cusp}}(\alpha)}{\beta(\alpha)}  
    \,,\nn \\
     A_{i}(\nu,\mu)&=-\int_{\alpha_{s}(\nu)}^{\alpha_{s}(\mu)} \mathrm{d}\alpha\, \frac{\gamma^{i}(\alpha)}{\beta(\alpha)}\,,\quad i=H,J,S  \,.
\end{align}

\subsection{Anomalous dimensions}
\label{appendix:ad}
Anomalous dimensions of the hard, jet and thrust soft functions up to three loops can be found in ref.~\cite{Becher:2006mr,Becher:2008cf}, and were obtained from~\cite{Moch:2005id,Moch:2004pa},\footnote{The convention for $\gamma^H$ in these references differs by a factor of 2 from ours. We follow ref.~\cite{Makris:2020ltr}.}
\begin{align}
\gamma_{0}^{H} &= -12 C_F\,, \nonumber
\\[2mm]
\gamma_{1}^{H} &=
2C_F^2\left(-3+4\pi^2-48\zeta_3\right)
+2 C_F C_A\left(
-\frac{961}{27}
-\frac{11\pi^2}{3}
+52\zeta_3
\right)
+2 C_F T_F n_f
\left(
\frac{260}{27}
+\frac{4\pi^2}{3}
\right), \nonumber
\\[2mm]
\gamma_{2}^{H} &=
2C_F^3\left(
-29-6\pi^2-\frac{16\pi^4}{5}
-136\zeta_3
+\frac{32\pi^2}{3}\zeta_3
+480\zeta_5
\right) \nonumber
\\
&\quad
+2 C_F^2 C_A\left(
-\frac{151}{2}
+\frac{410\pi^2}{9}
+\frac{494\pi^4}{135}
-\frac{1688}{3}\zeta_3
-\frac{16\pi^2}{3}\zeta_3
-240\zeta_5
\right) \nonumber
\\
&\quad
+ 2C_F C_A^2\left(
-\frac{139345}{1458}
-\frac{7163\pi^2}{243}
-\frac{83\pi^4}{45}
+\frac{7052}{9}\zeta_3
-\frac{88\pi^2}{9}\zeta_3
-272\zeta_5
\right) \nonumber
\\
&\quad
+2 C_F^2 T_F n_f\left(
\frac{5906}{27}
-\frac{52\pi^2}{9}
-\frac{56\pi^4}{27}
+\frac{1024}{9}\zeta_3
\right) \nonumber
\\
&\quad
+2 C_F C_A T_F n_f\left(
-\frac{34636}{729}
+\frac{5188\pi^2}{243}
+\frac{44\pi^4}{45}
-\frac{3856}{27}\zeta_3
\right) \nonumber
\\
&\quad
+2 C_F T_F^2 n_f^2\left(
\frac{19336}{729}
-\frac{80\pi^2}{27}
-\frac{64}{27}\zeta_3
\right),
\end{align}

\begin{align}
\gamma_{0}^{J} &= 6 C_F\,, \nonumber
\\[2mm]
\gamma_{1}^{J} &=
-2C_F^2\left(
-\frac{3}{2}+2\pi^2-24\zeta_3
\right)
-2 C_F C_A\left(
-\frac{1769}{54}
-\frac{11\pi^2}{9}
+40\zeta_3
\right)
-2 C_F T_F n_f\left(
\frac{242}{27}
+\frac{4\pi^2}{9}
\right), \nonumber
\\[2mm]
\gamma_{2}^{J} &=
-2C_F^3\left(
-\frac{29}{2}
-3\pi^2
-\frac{8\pi^4}{5}
-68\zeta_3
+\frac{16\pi^2}{3}\zeta_3
+240\zeta_5
\right) \nonumber
\\
&\quad
-2 C_F^2 C_A\left(
-\frac{151}{4}
+\frac{205\pi^2}{9}
+\frac{247\pi^4}{135}
-\frac{844}{3}\zeta_3
-\frac{8\pi^2}{3}\zeta_3
-120\zeta_5
\right) \nonumber
\\
&\quad
-2 C_F C_A^2\left(
-\frac{412907}{2916}
-\frac{419\pi^2}{243}
-\frac{19\pi^4}{10}
+\frac{5500}{9}\zeta_3
-\frac{88\pi^2}{9}\zeta_3
-232\zeta_5
\right) \nonumber
\\
&\quad
-2 C_F^2 T_F n_f\left(
\frac{4664}{27}
-\frac{32\pi^2}{9}
-\frac{164\pi^4}{135}
+\frac{208}{9}\zeta_3
\right) \nonumber
\\
&\quad
-2 C_F C_A T_F n_f\left(
-\frac{5476}{729}
+\frac{1180\pi^2}{243}
+\frac{46\pi^4}{45}
-\frac{2656}{27}\zeta_3
\right) \nonumber
\\
&\quad
-2 C_F T_F^2 n_f^2\left(
\frac{13828}{729}
-\frac{80\pi^2}{81}
-\frac{256}{27}\zeta_3
\right).
\end{align}
Due to consistency of the thrust factorization, the anomalous dimensions must satisfy 
\begin{align}
    \gamma^{S_{\mathrm{thr}}}=-\gamma^{H}-2\gamma^{J}\,,
    \label{eq:gSthr}
    \end{align}
such that
\begin{align}
\gamma_{0}^{S_{\mathrm{thr}}}&= 0\,, \nonumber
\\[2mm]
\gamma_{1}^{S_{\mathrm{thr}}} &
=-2 C_F C_A\left(
\frac{808}{27}
-\frac{11\pi^2}{9}
-28\zeta_3
\right)
-2 C_F T_F n_f
\left(
-\frac{224}{27}
+\frac{4\pi^2}{9}
\right), \nonumber
\\[2mm]
\gamma_{2}^{S_{\mathrm{thr}}} &=-2 C_F C_A^2\left(
\frac{273562}{1458}
-\frac{6325\pi^2}{243}
+\frac{88\pi^4}{45}
-\frac{3948}{9}\zeta_3
+\frac{88\pi^2}{9}\zeta_3
+192\zeta_5
\right) \nonumber
\\
&\quad
-2 C_F^2 T_F n_f\left(
-\frac{3368}{27}
+\frac{12\pi^2}{9}
+\frac{48\pi^4}{135}
+\frac{608}{9}\zeta_3
\right) \nonumber
\\
&\quad
-2 C_F C_A T_F n_f\left(
-\frac{23684}{729}
+\frac{2828\pi^2}{243}
-\frac{48\pi^4}{45}
+\frac{1456}{27}\zeta_3
\right) \nonumber
\\
&\quad
-2 C_F T_F^2 n_f^2\left(
-\frac{8320}{729}
-\frac{80\pi^2}{81}
+\frac{448}{27}\zeta_3
\right).
\end{align}
The cusp anomalous dimension is given in refs.~\cite{Moch:2004pa,Moch:2018wjh,Herzog:2018kwj,Henn:2019swt}.
The TMD anomalous dimension is fixed by consistency to
\begin{align}
\gamma^{V}=\frac{1}{2}\gamma^{H}
\,.\end{align}
We are left to obtain the anomalous dimension of the collinear-soft function, for which we use the consistency relation between the different anomalous dimensions of the functions present in the cross section in region 2:
\begin{align}
    \gamma^{H}+\gamma^{J}+\gamma^{S_{\mathrm{thr}}}-\gamma^{V}-\gamma^{{\mathscr{S}}}=0\,,
    \label{eq:gamma-sum}
\end{align}
From eqs.~(\ref{eq:gSthr})-(\ref{eq:gamma-sum}) we obtain
\begin{eqnarray}    \gamma^{{\mathscr{S}}}=\frac{\gamma^{S_{\mathrm{thr}}}}{2}\,.
\end{eqnarray}

\bibliography{bibFILE}
\end{document}